\documentclass[11pt,a4paper]{article}
\usepackage{amssymb,amsmath,amsthm,enumerate,graphicx,hyperref}
\usepackage{verbatim}
\usepackage{cite}
\usepackage[usenames]{color}
\definecolor{redch}{rgb}{0.8,0,0.2}

\usepackage{tikz}
\newcounter{nivell}
\newcommand{\nounivell}{%
  \addtocounter{nivell}{1}}
\newcommand{\nvl}{\value{nivell}}
\tikzstyle{hyb}=[rectangle,fill=green!50,draw,minimum size=5mm]
\tikzstyle{tre}=[circle,fill=green!50,draw,minimum size=5.5mm]
\newcommand{\etq}[1]{%
\draw (#1) node {$#1$};
}
\newcommand{\etqb}[2]{%
\draw (#1) node {$#1_{#2}$};
}
\usetikzlibrary{snakes}

\theoremstyle{plain}
\newtheorem{thm}{Theorem}
\newtheorem{prop}[thm]{Proposition}
\newtheorem{cor}[thm]{Corollary}
\newtheorem{lem}[thm]{Lemma}
\newenvironment{pf}{\begin{proof}}{\end{proof}}

\newcommand{\pathgr}{\!\rightsquigarrow\!{}}

\renewcommand{\leq}{\leqslant}
\renewcommand{\geq}{\geqslant}

\renewcommand{\ge}{\geqslant}
\newcommand{\NN}{\mathbb{N}}

\begin{document}
 
\title{Tripartitions do not always discriminate phylogenetic
networks\thanks{This work has been partially supported by
the Spanish CICYT project TIN2004-07925-C03-01 GRAMMARS, by Spanish
DGI projects MTM2006-07773 COMGRIO and MTM2006-15038-C02-01, and by
EU project INTAS IT 04-77-7178.}}

\author{\textbf{Gabriel Cardona}\\
Department of Mathematics\\
and Computer Science\\
University of the Balearic Islands\\
E-07122 Palma de Mallorca\\
Spain
\and
\textbf{Francesc Rossell\'o}\\
Research Institute of Health Science\\
University of the Balearic Islands\\
E-07122 Palma de Mallorca\\
Spain
\and \textbf{Gabriel Valiente}\\
Algorithms, Bioinformatics, Complexity\\
and Formal Methods Research Group\\
Technical University of Catalonia\\
E-08034 Barcelona\\
Spain
}
\maketitle

\begin{abstract}
Phylogenetic networks are a generalization of phylogenetic trees that
allow for the representation of non-treelike evolutionary events, like
recombination, hybridization, or lateral gene transfer.  In a recent
series of papers devoted to the study of reconstructibility of
phylogenetic networks, Moret, Nakhleh, Warnow and collaborators
introduced the so-called \emph{tripartition metric} for phylogenetic
networks.  In this paper we show that, in fact, this tripartition
metric does not satisfy the separation axiom of distances (zero distance means
isomorphism, or, in a more relaxed version, zero distance means
indistinguishability in some specific sense) in any of the subclasses of
phylogenetic networks where it is claimed to do so.  We also
present a subclass of phylogenetic networks whose members can be singled
out by means of their sets of tripartitions (or even clusters), and
hence where the 
latter can be used to define a meaningful metric.
\end{abstract}
\begin{quote}
\textbf{Keywords.} Phylogenetic networks, recombination, bipartitions,
tripartitions, tripartition metric, error metric
\end{quote}

\section{Introduction}

Phylogenetic trees have been used since the days of
Darwin~\cite{darwin:1837} to represent evolutionary histories of sets
of species under mutation.  Their popularity and prevalence have led
to the introduction of many methods to their reconstruction,
combination, and
comparison~\cite{dasgupta.ea:98,felsens:04,semple.steel:03}.  But, as
Doolittle pointed out about a decade ago~\cite{doolittle:99}, the
history of life cannot be properly represented as a tree.
Phylogenetic networks are used then as a generalization of
phylogenetic trees that allow for the representation of non-treelike
evolutionary events, like recombination, hybridization, or lateral
gene transfer~\cite{bandelt:94}.

The natural model for describing an evolutionary history is a directed
acyclic graph (DAG for short) representing the parent-child relation.
Phylogenetic trees are rooted DAGs where each node other than the root
(which represents the common ancestor of all individuals under
consideration, be them species or biomolecular sequences) has at most
one parent, from which it has been derived through mutation. Phylogenetic networks are rooted DAGs containing \emph{tree nodes},
which have only one parent and thus correspond to regular speciation
events, and \emph{hybrid nodes}, which have more than one parent and
thus correspond to hybrid speciation events. To such a DAG several
extra conditions have been added in the literature to provide a
realistic model of
recombination~\cite{strimmer.ea:2000,strimmer.ea:2001} or simply to
narrow the output space of reconstruction
algorithms~\cite{nakhleh.ea:bioinfo06,nakhleh.ea:bioinfo07}.

In a series of
papers~\cite{nakhleh.ea:TR0326,nakhleh.ea:tutpsb04,nakhleh.ea:TR0209,moret.ea:2004,%
nakhleh:phd04,nakhleh.ea:TR0406,nakhleh.ea:03psb} devoted to the study
of reconstructibility of phylogenetic networks, Moret, Nakhleh, Warnow
and collaborators introduced a method to compare a reconstructed
network and the true phylogeny, the so-called \emph{tripartition}, or
also \emph{error}, \emph{metric}.  This method is based on the
association, to each node $v$ of the network, of a tripartition of its
set of leaves into those that are strict descendants of $v$ (that is,
such that every path from the root to the leaf contains $v$), those
that are non-strict descendants of it (that is, that are descendants
but not strict descendants), and those that are not descendants of
$v$.  These tripartitions may be enriched with some extra information
like, for instance, the greatest number of hybrid nodes in a path from
$v$ to each leaf, or the sets of descendants of the parents of hybrid
nodes. Notice anyway that these tripartitions are a natural
generalization to non-tree networks of Bourque-Robinson-Foulds
bipartitions, which associate to each node of a phylogenetic tree the
partition of its leaves into descendant and
non-descendant ones. These bipartitions are used to define one of
the most popular distances for phylogenetic
trees~\cite{bourque:phd78,robinson.foulds:mb81}.

One of the key points in the definition of this tripartition metric is
the claim that the sets of tripartitions discriminate, up to isomorphism,
phylogenetic networks in a suitable subclass of them.  That is, that
if two phylogenetic networks $N$ and $N'$ in this subclass have the
same sets of tripartitions, then they are isomorphic.  This turns out
to be equivalent to the separation axiom for the tripartition metric
(zero distance means isomorphism).  In this paper, we provide
counterexamples showing that all claims made in this connection in
those papers are untrue, and thus that the tripartition metric does
not satisfy the separation axiom in any of the cases considered by the
authors, even in the more relaxed sense of~\cite{moret.ea:2004}, where
zero distance is simply claimed to be equivalent to equality modulo a
certain specific notion of indistinguishability.  Therefore, the
tripartition metric cannot be used in a meaningful way to compare
phylogenetic networks in the classes considered in the aforementioned
papers, as it cannot decide the  equality (or even indistinguishability) of networks.  Then, in the last section,
we show a slight variant of the class considered in one of these
papers where tripartitions, and even bipartitions in the sense of
Bourque-Robinson-Foulds, do define a metric.

\section{Notations on DAGs}
\label{sec:prl}

Let $N=(V,E)$ be a directed acyclic graph (DAG). 
We denote by $d_i(u)$ and $d_o(u)$ the in-degree and out-degree,
respectively, of a node $u\in V$. 

A node $v\in V$ is a \emph{leaf} if $d_o(v)=0$.  A node $v\in V$ is a
\emph{tree node} if $d_{i}(v)\leq 1$.  Such a tree node is a
\emph{root} if $d_i(v)=0$, and \emph{internal} if $d_i(v)=1$ and
$d_o(v)>0$.  A node $v\in V$ is \emph{hybrid} if $d_i(v)>1$.  We
denote by $V_L$, $V_{T}$, and $V_H$ the sets of leaves, of tree nodes,
and of hybrid nodes of $N$, respectively.  An arc $(u,v)\in E$ is a
\emph{tree arc} if its head $v$ is a tree node, and a \emph{network
arc} if $v$ is hybrid.

A \emph{clade} of a DAG $N$ is a subtree of $N$ with set of
nodes contained in $V_{T}$ and set of leaves contained in $V_{L}$.

A node $v\in V$ is a \emph{child} of $u\in V$ if $(u,v)\in E$; we also
say that $u$ is a \emph{parent} of $v$.  
All children of the same
node are said to be \emph{siblings} of each other.  The \emph{tree
children} of a node $u$ are its children that are tree nodes.

Let $S$ be any finite set of \emph{labels}.  We say that the DAG $N$
is \emph{labeled in $S$} when its leaves are bijectively labeled by
elements of $S$.  Two DAGs $N,N'$ labeled in $S$ are
\emph{isomorphic}, in symbols $N\cong N'$, when they are isomorphic as
directed graphs and the isomorphism preserves the leaves' labels.
 
We shall always assume, usually without any further notice,
that the DAGs appearing in this paper are labeled in some set $S$, and
we shall always identify, usually without any further notice either,
each leaf of a DAG with its label in~$S$.

A \emph{path} on a DAG $N=(V,E)$ is a sequence of nodes
$(v_0,v_1,\dots,v_k)$ such that $(v_{i-1},v_{i})\in E$ for all
$i=1,\dots,k$; such a path is a \emph{cycle} if $v_{k}=v_0$.  We
call $v_0$ the \emph{origin} of the path, $v_{1},\ldots,v_{k-1}$ its
\emph{intermediate nodes}, and $v_{k}$ its \emph{end}.  The
\emph{length} of the path $(v_0,v_1,\dots,v_k)$ is $k$, and it is
\emph{non-trivial} if $k\ge 1$.  We denote by $u\pathgr v$ any path with
origin $u$ and end $v$.  

The relation $\geq$ on $V$ defined by
$$
u\geq v \iff \mbox{there exists a path $u\pathgr v$}
$$
is a partial order, called the \emph{path ordering} on $N$.  Whenever
$u\geq v$, we shall say that $v$ is a \emph{descendant} of $u$ and
also that $u$ is an \emph{ancestor} of $v$.

\section{Tripartitions}
\label{sec:trip}

Let $N=(V,E)$ be any DAG labeled in $S$. For every node $u\in V$:
\begin{itemize}
\item Let $C(u)\subseteq S$ be the set of leaves that are
descendants of $u$. We call $C(u)$ the \emph{cluster} of $u$.

\item Let $A(u)\subseteq C(u)$ be the set of leaves that are \emph{strict
descendants} of $u$: that is, those leaves $s$ such that every path
from a root of $N$ to $s$ contains the node $u$.  We call $A(u)$ the
\emph{strict cluster} of $u$.

\item Let $B(u)\subseteq C(u)$ be the set $C(u)\setminus A(u)$ of leaves
that are \emph{non-strict descendants} of $u$: those leaves $s$ that
are descendants of $u$, but for which there exists some path from a
root to $s$ not containing the node $u$.

\item Let $C^c(u)\subseteq S$ be the set $V_{L} \setminus C(u)$ of
leaves that are not descendants of $u$.
\end{itemize}

A \emph{phylogenetic tree} on a set $S$ of taxa is a rooted tree with
its leaves labeled bijectively in $S$, i.e., a rooted DAG labeled in
$S$ without hybrid nodes.  Notice that, in a phylogenetic tree,
$C(u)=A(u)$ and $B(u)=\emptyset$ for every node $u$.  This property
actually characterizes phylogenetic trees among all rooted DAGs.

Every arc $e=(u,v)$ of a phylogenetic tree $T=(V,E)$ on $S$ defines a
\emph{bipartition} of $S$ 
$$
\pi^{(T)}(e)=(C(v),C^c(v)).
$$
Let $\pi(T)$ denote the set of all these bipartitions:
$$
\pi(T)=\{\pi^{(T)}(e)\mid e\in E\}.
$$

The \emph{bipartition distance}
\cite{bourque:phd78,robinson.foulds:mb81} between two phylogenetic
trees $T$ and $T'$ on the same set $S$ of taxa is defined as
$$
d_{\pi}(T,T')=\frac{1}{2}\big(|\pi(T)\setminus \pi(T')|+|\pi(T')\setminus
\pi(T)|\big).
$$

The bipartition distance is a true distance for phylogenetic trees, in
the sense that it satisfies the axioms of distances up to isomorphisms:
for every phylogenetic trees $T,T',T''$ on the same set $S$ of taxa,
\begin{enumerate}[(a)]
\item \emph{Non-negativity}: $d_{\pi}(T,T')\geq 0$
\item \emph{Separation}: $d_{\pi}(T,T')=0$ if and only if $T\cong T'$
\item \emph{Symmetry}: $d_{\pi}(T,T')=d_{\pi}(T',T)$
\item \emph{Triangle inequality}: $d_{\pi}(T,T')\leq d_{\pi}(T,T'')+d_{\pi}(T'',T')$
\end{enumerate}

Phylogenetic networks \cite{bandelt:94} are usually defined as a
subclass of DAGs that extend phylogenetic trees by allowing the
existence of hybrid nodes representing recombination or lateral gene
transfer events.  In a
non-tree DAG $N$ it still makes sense to consider bipartitions $\pi$
associated to arcs $e=(u,v)$,
$$
\pi^{(N)}(e)=(C(v),C^c(v)),
$$
and then to define
$$
\pi(N)=\{\pi^{(N)}(e)\mid e\in E\}.
$$
and
$$
d_{\pi}(N,N')=\frac{1}{2}\big(|\pi(N)\setminus \pi(N')|+|\pi(N')\setminus
\pi(N)|\big).
$$
As we shall see in Section~\ref{sec:treechild}, there even exist
subclasses of non-tree DAGs
where $d_{\pi}$ is a distance.

But, to distinguish between strict and non-strict descendants gives more
information about the topological relations between the hybrid nodes.
This is the reason why, in the series of papers 
\cite{nakhleh.ea:TR0326,nakhleh.ea:tutpsb04,nakhleh.ea:TR0209,moret.ea:2004,%
nakhleh:phd04,nakhleh.ea:TR0406,nakhleh.ea:03psb}, Moret, Nakhleh,
Warnow, and collaborators associate to each arc $e=(u,v)$ of a DAG $N$
labeled in $S$, the \emph{tripartition} of $S$
$$
\theta^{(N)}(e)=(A(v), B(v), C^c(v)).
$$
As an extra piece of information about the topology of the DAG, the
leaves $s$ in $A(v)$ and $B(v)$ can be weighted with the maximum number
of hybrid nodes contained in a path from $v$ to $s$ (including $v$ and
$s$ themselves).  Therefore, we can distinguish between:

\begin{itemize}
\item The (\emph{unweighted}) \emph{tripartition} $\theta^{(N)}(e)$
\cite{nakhleh.ea:TR0406,nakhleh:phd04,moret.ea:2004}.

\item The \emph{$B$-weighted tripartition} $\theta^{(N)}_{B}(e)$, where the 
elements of $B(v)$ are weighted as indicated \cite{nakhleh.ea:TR0209}.

\item The \emph{$AB$-weighted tripartition} $\theta^{(N)}_{AB}(e)$, where the
elements of both $A(v)$ and $B(v)$ are weighted as indicated
\cite{nakhleh.ea:TR0326,nakhleh.ea:tutpsb04}.

\end{itemize}

For every type of tripartition
$\Upsilon=\theta,\theta_{B},\theta_{AB}$, let us denote by
$\Upsilon(N)$ the set of all these tripartitions of arcs of $N$:
$$
\Upsilon(N)=\{\Upsilon^{(N)}(e)\mid e\in E\}.
$$
The \emph{tripartition distance relative to $\Upsilon$} between two DAGs
$N_{1}=(V_{1},E_{1})$ and $N_{2}=(V_{2},E_{2})$ labeled in the same
set $S$ can be defined then, by analogy with the bipartition distance,
by
$$
d_{\Upsilon}(N_{1},N_{2})=\frac{1}{2}\big(| \Upsilon(N_{1})\setminus
\Upsilon(N_{2})|+| \Upsilon(N_{2})\setminus
\Upsilon(N_{1})|\big),
$$

It is obvious that $d_{\Upsilon}$ always satisfies the non-negativity,
symmetry and triangle inequality axioms of distances on the
class of all DAGs labeled in a fixed set $S$.  As far as the
separation axiom goes, notice that $d_{\Upsilon}(N_{1},N_{2})=0$ if
and only if $\Upsilon(N_{1})= \Upsilon(N_{2})$.  Therefore,
$d_{\Upsilon}$ satisfies the separation axiom on a certain subclass of
DAGs if and only if $\Upsilon(N)$ characterizes $N$ up to isomorphism
among all DAGs in this subclass.  When these equivalent conditions
hold, we shall say that the tripartition $\Upsilon$ \emph{satisfies
the separation property} on the subclass of DAGs under consideration.

Notice that, for every DAGs $N$ and $N'$,
$$
\theta_{AB}(N)=\theta_{AB}(N')\Longrightarrow
\theta_{B}(N)=\theta_{B}(N')\Longrightarrow
\theta(N)=\theta(N').
$$
Then, the separation property for $\theta$ implies the separation
property for $\theta_{B}$, and the latter implies the separation
property for $\theta_{AB}$.

In the papers by Moret, Nakhleh, Warnow, et al mentioned above, it is
claimed that some of these tripartitions $\Upsilon$ (and even further
refinements of them) satisfy the separation property on some specific
subclasses of DAGs.  In the following sections we show that all these
claims are incorrect, and then in Section \ref{sec:treechild} we show
a subclass of phylogenetic networks where $\theta$, an even the
bipartition $\pi$, satisfy the separation property.

To end this section, we want to mention that in the aforementioned
papers, the $\Upsilon$-tripartition distance is actually not defined as above,
but in a normalized version, which the authors call \emph{error
metric}:
$$
m_{\Upsilon}(N_{1},N_{2})=\frac{1}{2}\Bigl(\frac{| \Upsilon(N_{1})\setminus
\Upsilon(N_{2})|}{|E_{1}|}+\frac{| \Upsilon(N_{2})\setminus
\Upsilon(N_{1})|}{|E_{2}|}\Bigr).
$$
This is the function claimed to be a distance on some subclasses of DAGs
in those papers.  It is straightforward to notice that $m_{\Upsilon}$
satisfies the separation axiom of distances on a subclass of DAGs if and
only if $d_{\Upsilon}$ does so, and therefore the counterexamples in
the next sections also entail that $m_{\Upsilon}$ neither satisfies
the separation axiom  when claimed.

But, contrary to what happens
with   $d_{\Upsilon}$ and against what is claimed in
several of the papers under review (see, for instance, the proof of
\cite[Thm.~3]{moret.ea:2004}), this normalized version $m_{\Upsilon}$
need not satisfy the triangle inequality either, even on the subclasses
of DAGs considered in those papers.  For one reason, the failure of
the separation property allows the existence of DAGs $N=(V,E)$ and
$N'=(V',E')$ such that $\Upsilon(N)=\Upsilon(N')$ but, say,
$|E|<|E'|$.  Then, given any other DAG $N_{0}=(V_{0},E_{0})$,
$$
\begin{array}{rl}
m_{\Upsilon}(N_{0},N) & \displaystyle =\frac{1}{2}\Bigl(\frac{| \Upsilon(N_{0})\setminus
\Upsilon(N)|}{|E_{0}|}+\frac{| \Upsilon(N)\setminus
\Upsilon(N_{0})|}{|E|}\Bigr)\\[2ex]
m_{\Upsilon}(N_{0},N') & \displaystyle  =\frac{1}{2}\Bigl(\frac{| \Upsilon(N_{0})\setminus
\Upsilon(N')|}{|E_{0}|}+\frac{| \Upsilon(N')\setminus
\Upsilon(N_{0})|}{|E'|}\Bigr)\\[2ex]
& \displaystyle  =\frac{1}{2}\Bigl(\frac{| \Upsilon(N_{0})\setminus
\Upsilon(N)|}{|E_{0}|}+\frac{| \Upsilon(N)\setminus
\Upsilon(N_{0})|}{|E'|}\Bigr)\\[2ex]
m_{\Upsilon}(N',N) & \displaystyle  =\frac{1}{2}\Bigl(\frac{| \Upsilon(N')\setminus
\Upsilon(N)|}{|E'|}+\frac{| \Upsilon(N)\setminus
\Upsilon(N')|}{|E|}\Bigr)=0
\end{array}
$$
and then $|E|<|E'|$ implies that
$$
\begin{array}{rl}
m_{\Upsilon}(N_{0},N) & \displaystyle =\frac{1}{2}\Bigl(\frac{| \Upsilon(N_{0})\setminus
\Upsilon(N)|}{|E_{0}|}+\frac{| \Upsilon(N)\setminus
\Upsilon(N_{0})|}{|E|}\Bigr)\\[2ex]
& \displaystyle  >\frac{1}{2}\Bigl(\frac{| \Upsilon(N_{0})\setminus
\Upsilon(N)|}{|E_{0}|}+\frac{| \Upsilon(N)\setminus
\Upsilon(N_{0})|}{|E'|}\Bigr)+0\\[2ex]
& =m_{\Upsilon}(N_{0},N')+m_{\Upsilon}(N',N).
\end{array}
$$
For two specific $N=(V,E)$ and $N'=(V',E')$ such that
$\Upsilon(N)=\Upsilon(N')$ but $|E|\neq |E'|$, precisely in the
context of Thm.~3 in \cite{moret.ea:2004}, see the DAGs $N_{9}$ and
$N_{10}$ depicted in Fig.~\ref{fig:contr6} in the Appendix\footnote{To make easier reading this paper, we
gather all large figures depicting phylogenetic networks, as well as
the tables giving their tripartitions, in an Appendix at the end of the
paper.}.  They are reduced reconstructible
phylogenetic networks (see Section~\ref{sec:ieee} for the precise
meaning of these words), the subclass where that theorem claims that
$m_{\theta}$ satisfies the triangle inequality, and they have the same
$AB$-weighted tripartitions (see Table~\ref{tbl:contr7.2} also
in the Appendix) but
different numbers of arcs.  This shows that $m_{\theta_{AB}}$ does not
satisfy the triangle inequality on the subclass of reduced
reconstructible phylogenetic networks.

Anyway, the failure of the triangle inequality is easily solved for
instance using $d_\Upsilon$ instead of $m_\Upsilon$.  The failure of
the separation axiom is deeper, as it reflects the impossibility of
discriminating the phylogenetic networks under consideration using only
information on their tripartitions.

\section{The first error metric}
\label{sec:primo}

The error metric for phylogenetic networks is introduced by
Moret, Nakhleh, and Warnow in the Technical Report
\cite{nakhleh.ea:TR0209}.  In it, a \emph{phylogenetic network} on a
set $S$ of taxa is defined as a rooted DAG $N$ labeled in $S$
satisfying the following conditions:
\begin{enumerate}[(\thesection.1)]
\item The in-degree and out-degree of each node is 0, 1, or 2, and no
node has its in-degree equal to its out-degree.

\item If a node has two children, at least one of them is a tree node.

\item \emph{Weak time consistency:}
\begin{enumerate}

\item[(\thesection.3.a)] If $u_{1}$ and $u_{2}$ are the parents of a
hybrid node $u$, then there do not exist paths $u_{1}\pathgr u_{2}$ or
$u_{2}\pathgr u_{1}$.

\item[(\thesection.3.b)] If $u_{1}$ and $u_{2}$ are the parents of a
hybrid node $u$, and $v_{1}$ and $v_{2}$ are the parents of a hybrid
node $v$, and there exists a path $u_{1}\pathgr v_{1}$, then there do
not exist paths $v_{2}\pathgr u_{1}$ or $v_{2}\pathgr u_{2}$.
\end{enumerate}
\end{enumerate}
Notice that in these phylogenetic networks, a hybrid node can have a
hybrid child. This would correspond to a hybrid node that hybridizes
before undergoing a speciation event, a scenario that, the authors say,
``almost never arises in reality.''

Condition (4.3) is a first, weak version of a constant property of
Nakhleh-Warnow-{et al}'s phylogenetic networks: time consistency.
Roughly speaking, this property aims at assigning times to the nodes
of the network in a way that strictly increases on tree arcs and so
that the parents of a hybrid node coexist in time; we shall discuss
 this topic further in the next section.  In the weak
version recalled in this section, and which does not entail in general
such a timing (cf.~Prop.~\ref{prop:timing} in the next section), this
restriction is simply imposed by not allowing a node to hybridize with
its descendants, and by forbidding the ancestors of a parent of a
hybrid node to hybridize with the descendants of the other parent.

A phylogenetic network is said to be of \emph{class I} when each
hybrid node has at least one parent that is a tree node.

Then, it is claimed in \cite[Thm.~4]{nakhleh.ea:TR0209} that
$\theta_B$ satisfies the separation property on the subclass of all
class I phylogenetic networks.  This claim is untrue,
because there exist pairs of non-isomorphic class I phylogenetic
networks with the same sets of $B$-weighted (even $AB$-weighted)
tripartitions.

Consider for instance the phylogenetic networks $N_{1}$ and $N_{2}$
labeled in $\{1,\!\ldots,\!  5\}$ depicted in Fig.~\ref{fig:contr1} in
the Appendix.  It 
is easy to check that these two DAGs satisfy conditions (4.1) to (4.3)
above and that they are of class I. Now, Table \ref{tbl:contr1}
displays the $AB$-weighted tripartitions of these networks induced by
their arcs.  A simple inspection of this table shows that
$\theta_{AB}(N_{1})=\theta_{AB}(N_{2})$.  But it is clear that
$N_{1}\not\cong N_{2}$.

It is interesting to point out that $\theta_B$, and even $\theta$,
satisfy the separation property if we moreover forbid the
``improbable'' event of two consecutive hybridizations: i.e., if we
impose not only that some child of a tree node is a tree node
(condition (4.2)), but also that the only child of an internal hybrid
node is a tree node, a condition that would be imposed in later
versions (see conditions (5.2), (6.2), and (8.2)).  We devote
Section~\ref{sec:treechild} to prove this fact.

\section{The second version: introducing strong time consistency}
\label{sec:second}

A new version of the tripartition metric is presented in the Technical Report
\cite{nakhleh.ea:TR0326}.  This is the metric used, for
instance, in the paper \cite{nakhleh.ea:03psb}.  The main difference between
the proposal in this new Technical Report and the previous one
\cite{nakhleh.ea:TR0209} is the refinement of the
notion of phylogenetic network, by distinguishing between \emph{model}
and \emph{reconstructible} phylogenetic networks and
 strengthening the time compatibility and class I
conditions.

In \cite{nakhleh.ea:TR0326}, a \emph{model phylogenetic network} on a
set $S$ of taxa is defined as a rooted DAG $N$ labeled in $S$
satisfying the following conditions:
\begin{enumerate}[(\thesection.1)]
\item The root and all internal tree nodes have out-degree 2.  All
hybrid nodes have out-degree 1, and they can only have in-degree 2
(allo-polyploid hybrid nodes) or 1 (auto-polyploid hybrid
nodes).\footnote{In our classification of nodes in DAGs in
Section~\ref{sec:prl}, these auto-polyploid hybrid nodes, with
in-degree and out-degree equal to 1, would actually be considered tree
nodes.}

\item The child of a hybrid node is always a tree node.

\item \emph{Strong time consistency:} Let $x,y$ be any two nodes for which
there exists a sequence of nodes $(v_0,v_{1},\ldots,v_{k})$ with
$v_0=x$ and $v_{k}=y$ such that:
\begin{itemize}
    \item for every $i=0,\ldots,k-1$, either $(v_{i},v_{i+1})$ is an arc of $N$,
or $(v_{i+1},v_{i})$ is a network arc of $N$;

\item at least one pair $(v_{i},v_{i+1})$ is a tree arc of $N$;
\end{itemize}
(that is, $(v_0,\ldots,v_{k})$ is a path from $x$ to $y$ containing
some tree arc of $N$, in the graph $N^{*}$ obtained from $N$ by adding
the inverses of all network arcs).  Then, $x$ and $y$ cannot have a
hybrid child in common.
\end{enumerate}

This new notion of time consistency generalizes the one given
in the previous version, and, as we shall see in a minute, it captures
the notion of timing mentioned therein.  This timing is given by a
\emph{temporal representation} in the sense of Baroni, Semple and
Steel \cite{BSS:06}: a mapping $\tau :V\to \NN$ such that:
\begin{enumerate}[(a)]
\item if $r$ is the root of $N$, then $\tau(r)=0$;
\item if $(u,v)\in E_{T}$, then $\tau(u)<\tau(v)$;
\item if $(u,v)\in E_{N}$, then $\tau(u)=\tau(v)$.
\end{enumerate}
Baroni, Semple and
Steel prove in \textsl{loc.  cit.} the equivalence between the
existence of such a temporal representation and the fact that a  certain
quotient graph (essentially obtained by identifying hybrid nodes with
their parents) of the network is acyclic.  Since none of the papers on
tripartitions we are discussing provides a formal proof of the fact
that condition (\thesection.3) above is equivalent to the existence of
a temporal representation, and for the sake of completeness, we
provide such a proof here.

\begin{prop}
    \label{prop:timing}
Let $N=(V,E)$ be a rooted DAG, let $E_{T}$ and $E_{N}$ be its sets of
tree and network arcs, respectively, and let $N^{*}=(V,E^{*})$ be the
directed graph with the same set $V$ of nodes as $N$ and set of arcs
$E^{*}=E\cup E_{N}^{-1}$.  The following conditions are equivalent:
\begin{enumerate}[(i)]
\item $N$ is  strongly time consistent.

\item $N^{*}$ does not have any cycle containing some tree arc of
$N$.

\item $N$ admits a temporal representation.
\end{enumerate}

\end{prop}

\begin{pf}
(i)$\Longrightarrow$(ii)
To begin with, notice that if  $N^{*}$ has cycles containing tree arcs of
$N$, then it has a minimal\footnote{By a \emph{minimal 
cycle} $(v_0,v_{1},\ldots,v_{k},v_0)$ we mean a cycle such that
the nodes $v_0,v_1,\ldots,v_{k}$ are pairwise different.}
such cycle.  Indeed, if $N^{*}$ has cycles containing tree arcs, then it
will contain one of shortest length, say
$$
(v_0,v_{1},\ldots,v_{k},v_0).
$$
If it is not minimal, then $v_{i}=v_{j}$ for some
$0\leq i<j\leq k$ (actually $j-i\geq 2$, because $N$ does not contain 
loops), and hence we have two strictly shorter cycles in $N^{*}$,
$$
(v_{i},v_{i+1},\ldots,v_{j-1},v_{j}=v_{i})\mbox{ and }
(v_0,v_{1},\ldots,v_{i-1},v_{i}=v_{j},v_{j+1},\ldots,v_{k},v_0),
$$
and at least one of them contains the tree arc that belonged to the
original cycle, which leads to a contradiction.

Assume now that $N$ satisfies the strong time consistency condition
and that $N^*$ has a minimal cycle
$$
(v_0,v_{1},\ldots,v_{k},v_0)
$$
 containing some tree arc of $N$: without any loss
of generality, we shall assume that $(v_{k},v_0)$ is a tree arc of
$N$, and in particular that $v_0$ is a tree node.  

Since $N$ is acyclic, this cycle must contain some arc in
$E_{N}^{-1}$.  Let $i\in \{1,\ldots,k-1\}$ be the first index such
that $(v_{i},v_{i+1})\in E_{N}^{-1}$: in particular, $v_{i+1}$ is one
of the parents in $N$ of the hybrid node $v_{i}$ (notice that $i\neq
0$, because $v_0$ is a tree node of $N$, and that $i\neq k$ because
$(v_{k},v_0)$ is a tree arc of $N$).  Then the arc $(v_{i-1},v_{i})$
is a network arc of $N$, and since the considered cycle is minimal,
$v_{i-1}$ must be different from $v_{i+1}$.

But in this case the sequence of nodes
$$
(v_{i+1},\ldots,v_{k},v_0,v_{1},\ldots,v_{i-1})
$$
is a path in $N^{*}$ containing at least one tree arc and connecting two
parents of a hybrid node of $N$, which contradicts the strong time
consistency condition.  This shows that $N^{*}$ cannot contain any
minimal cycle containing some tree arc of $N$.

(ii)$\Longrightarrow$(iii)
If  $N^{*}$ does not have any cycle containing some tree arc of
$N$, then the mapping
$$
\tau: V \to \NN
$$
that sends each $v\in V$ to the maximum number of tree arcs in a path
from the root $r$ to $v$ in $N^{*}$ is well defined, and it clearly satisfies
conditions (a) to (c) in the statement.

(iii)$\Longrightarrow$(i) Assume that a mapping $\tau$ as described in
(iii) exists.  Then, if $(v_0,\ldots,v_{k})$ is a path in $N^{*}$
containing some tree arc of $N$, we have that
$\tau(v_0)<\tau(v_{k})$, and then $v_0$ and $v_{k}$ cannot be the
parents of a hybrid node $u$, because the latter would imply that
$\tau(v_0)=\tau(u)=\tau(v_{k})$.  \qed
\end{pf}

In \emph{reconstructible phylogenetic networks} the previous
conditions are relaxed: tree nodes can have any out-degree greater than
1; hybrid nodes can have any in-degree greater than 1 and any out-degree
greater than 0 (in particular, auto-polyploid hybrid nodes are
forbidden); a hybrid node can have hybrid children; and the strong
time consistency need not hold any longer.

Now, two nodes $u,v$ of a (model or reconstructible) phylogenetic
network are said to be \emph{convergent} when they satisfy the
following condition:
\begin{quote}
For every leaf $s$ and for every $k\geq 0$, there exists a path
$u\pathgr s$ containing $k$ hybrid nodes if and only if there exists a
path $v\pathgr s$ containing $k$ hybrid nodes.
\end{quote}
A phylogenetic network is said to be \emph{of class I} if it does not 
contain any pair of convergent nodes. 

\begin{figure}[htb]
\begin{center}
\begin{tikzpicture}[thick,>=stealth,scale=0.8]
\draw (1,0) node[tre] (r) {}; \etq r
\draw (0,-1) node[tre] (c) {}; \etq c
\draw (2,-1) node[tre] (d) {}; \etq d
\draw (1,-2) node[hyb] (A) {}; \etq A
\draw (0,-3) node[tre] (b) {}; \etq b 
\draw (3,-3) node[hyb] (B) {}; \etq B
\draw (5,-3) node[tre] (e) {}; \etq e
\draw (0,-4) node[tre] (a) {}; \etq a
\draw (4,-4) node[hyb] (C) {}; \etq C
\draw (5,-4) node[tre] (f) {}; \etq f
\draw (4,-5) node[tre] (g) {}; \etq g
\draw (3,-6) node[hyb] (D) {}; \etq D
\draw (2,-7) node[tre] (h) {}; \etq h
\draw (1,-8) node[hyb] (E) {}; \etq E
\draw (0,-9) node[tre] (1) {}; \etq 1
\draw (1,-9) node[tre] (2) {}; \etq 2
\draw (2,-9) node[tre] (3) {}; \etq 3
\draw (4,-9) node[tre] (4) {}; \etq 4
\draw (5,-9) node[tre] (5) {}; \etq 5
\draw [->](r)--(c);
\draw [->](r)--(d);
\draw [->](c)--(A);
\draw [->](c)--(b);
\draw [->](d)--(A);
\draw [->](d)--(e);
\draw [->](A)--(E);
\draw [->](b)--(a);
\draw [->](b)--(B);
\draw [->](B)--(D);
\draw [->](e)--(B);
\draw [->](e)--(f);
\draw [->](a)--(1);
\draw [->](a)--(C);
\draw [->](f)--(C);
\draw [->](C)--(g);
\draw [->](g)--(D);
\draw [->](g)--(4);
\draw [->](D)--(h);
\draw [->](h)--(E);
\draw [->](h)--(3);
\draw [->](E)--(2);
\draw [->](f)--(5);
\end{tikzpicture}
\qquad
\begin{tikzpicture}[thick,>=stealth,scale=0.8]
\draw (1,0) node[tre] (r) {}; \etq r
\draw (0,-1) node[tre] (c) {}; \etq c
\draw (2,-1) node[tre] (d) {}; \etq d
\draw (1,-2) node[hyb] (A) {}; \etq A
\draw (0,-3) node[tre] (b) {}; \etq b 
\draw (3,-3) node[hyb] (B) {}; \etq B
\draw (5,-3) node[tre] (e) {}; \etq e
\draw (0,-4) node[tre] (a) {}; \etq a
\draw (4,-4) node[hyb] (C) {}; \etq C
\draw (5,-4) node[tre] (f) {}; \etq f
\draw (4,-5) node[tre] (g) {}; \etq g
\draw (3,-6) node[hyb] (D) {}; \etq D
\draw (2,-7) node[tre] (h) {}; \etq h
\draw (1,-8) node[hyb] (E) {}; \etq E
\draw (0,-9) node[tre] (1) {}; \etq 1
\draw (1,-9) node[tre] (2) {}; \etq 2
\draw (2,-9) node[tre] (3) {}; \etq 3
\draw (4,-9) node[tre] (4) {}; \etq 4
\draw (5,-9) node[tre] (5) {}; \etq 5
\draw [->](r)--(c);
\draw [->](r)--(d);
\draw [->](c)--(A);
\draw [->](c)--(b);
\draw [->](d)--(A);
\draw [->](d)--(e);
\draw [red,->](A)--(D);
\draw [->](b)--(a);
\draw [->](b)--(B);
\draw [red,->](B)--(E);
\draw [->](e)--(B);
\draw [->](e)--(f);
\draw [->](a)--(1);
\draw [->](a)--(C);
\draw [->](f)--(C);
\draw [->](C)--(g);
\draw [->](g)--(D);
\draw [->](g)--(4);
\draw [->](D)--(h);
\draw [->](h)--(E);
\draw [->](h)--(3);
\draw [->](E)--(2);
\draw [->](f)--(5);
\end{tikzpicture}
\end{center}
\caption{\label{fig:contr1} The networks $N_{1}$ (left) and $N_{2}$
(right)}
\end{figure}
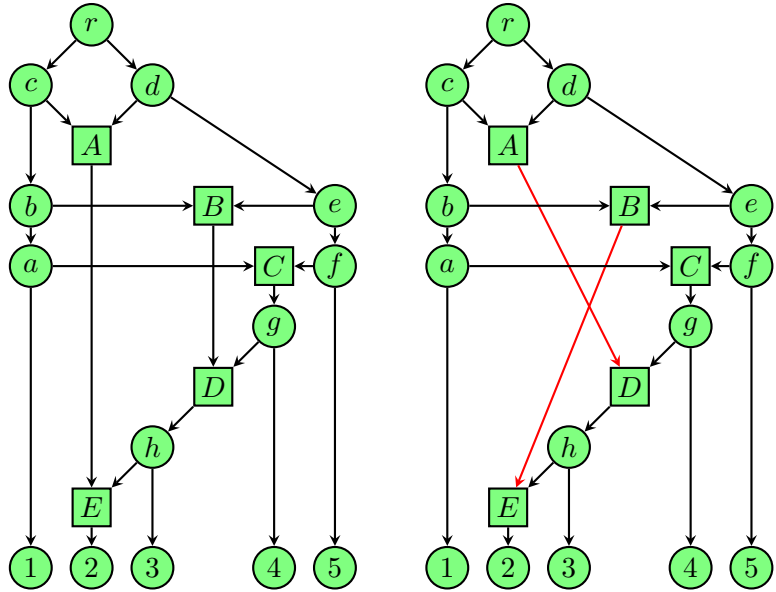

Then, it is claimed in \cite[Thm.~4]{nakhleh.ea:TR0326} that
$\theta_{AB}$ satisfies the separation property on this new 
class I of phylogenetic networks.  It is again incorrect, because there
still exist pairs of non-isomorphic class I model phylogenetic
networks with $AB$-weighted error rate 0.  Consider for instance the
model phylogenetic networks $N_{3}$ and $N_{4}$ labeled in
$\{1,\ldots,13\}$ depicted in Fig.~\ref{fig:contr2}  in the Appendix, which are
suitable modifications of those given in Fig.~\ref{fig:contr1} to cope
with the new restrictions on model phylogenetic networks (much simpler
examples exist involving reconstructible networks containing, for
instance, out-degree 3 tree nodes: consider the networks $N_{9}$ and
$N_{10}$ shown in Fig.~\ref{fig:contr6}).

These networks have no pair of convergent nodes (as it can be checked
in Table \ref{tbl:contr2}) and they are clearly non-isomorphic.  Now,
Table \ref{tbl:contr2} displays the tripartitions of these networks
induced by their arcs and, again, a simple inspection of this table
shows that $\theta_{AB}(N_{3})=\theta_{AB}(N_{4})$.

\section{The third version: introducing the reticulation scenarios}

In the Technical Reports \cite{nakhleh.ea:tutpsb04,nakhleh.ea:TR0406} 
the authors present a third version of the tripartition metric, now including
a substantial change in the definition of the metric itself.

The definition of phylogenetic network in these papers is that of
model phylogenetic network in the previous section with a small
modification in the time compatibility condition.  More specifically,
a \emph{phylogenetic network} on a set $S$ of labels is a rooted DAG
$N$ labeled in $S$ satisfying the following conditions:
\begin{enumerate}[(\thesection.1)]
\item The root and all internal tree nodes have out-degree 2.  All
hybrid nodes have out-degree 1, and they can only have in-degree 2
or (in \cite{nakhleh.ea:tutpsb04}) 1.
    
\item The child of a hybrid node is always a tree node.

\item \emph{Time consistency:} Let $x,y$ be any two nodes for which
there exists a sequence of paths $(P_0,P_{1},\ldots,P_{k})$ in $N$
such that
\begin{itemize}
    \item $x$ is the origin of $P_0$ and $y$ is the end of $P_{k}$;
    \item each path contains some tree arc;
    \item for every $i=0,\ldots,k-1$, the end of $P_{i}$ and the
    origin of $P_{i+1}$ are the parents of a hybrid node.
\end{itemize}
Then, $x$ and $y$ cannot have a hybrid child in common.
\end{enumerate}

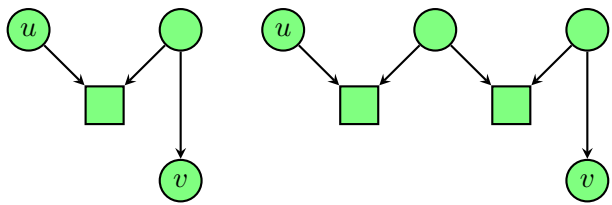
\begin{figure}[htb]
\begin{center}
\begin{tikzpicture}[thick,>=stealth]
\draw (0,0) node[tre] (u) {}; \etq u
\draw (2,0) node[tre] (x) {}; 
\draw (1,-1) node[hyb] (y) {}; 
\draw (2,-2) node[tre] (v) {}; \etq v
\draw [->](u)--(y);
\draw [->](x)--(y);
\draw [->](x)--(v);
\end{tikzpicture}\qquad
\begin{tikzpicture}[thick,>=stealth]
\draw (0,0) node[tre] (u) {}; \etq u
\draw (2,0) node[tre] (x) {}; 
\draw (1,-1) node[hyb] (y) {}; 
\draw (4,0) node[tre] (xb) {}; 
\draw (3,-1) node[hyb] (yb) {}; 
\draw (4,-2) node[tre] (v) {}; \etq v
\draw [->](u)--(y);
\draw [->](x)--(yb);
\draw [->](x)--(y);
\draw [->](xb)--(yb);
\draw [->](xb)--(v);
\end{tikzpicture}
\end{center}
\caption{\label{fig:tc-nostc} Time compatibility allows the nodes $u$
and $v$ in these graphs to hybridize, while strong time compatibility
does not.}
\end{figure}

This time consistency condition (6.3) is actually weaker than the
strong time consistency (5.3).  For instance, Fig.~\ref{fig:tc-nostc}
shows two situations where condition (6.3) allows the nodes $u$ and
$v$ to hybridize, while condition (5.3) forbids it.  Notice
nevertheless that, under condition (6.2), time consistency (6.3)
becomes equivalent to strong time consistency (5.3) if we simply ask
\emph{at least one of the paths $P_i$} (instead of all of
them) to contain some tree arc.

In these papers the authors do not consider any class I of phylogenetic
networks, but instead they refine the error rate by adding some extra
information to tripartitions induced by network arcs.  Namely, they
define the \emph{reticulation scenario} $RS(v)$ of a hybrid node $v$ with
parents $u_{1},u_{2}$ as the set of clusters of its parents:
$$
RS(v)=\{C(u_{1}),C(u_{2})\}.
$$
Then, the data the authors\footnote{Actually, the tripartitions they
use are $AB$-weighted in \cite{nakhleh.ea:tutpsb04} and unweighted in
\cite{nakhleh.ea:TR0406}.  For the sake of generality, in this section
we shall use $\theta_{AB}$.} consider on arcs are:
\begin{itemize}
\item If $e$ is a tree arc, then $\Psi^{(N)}(n)=\theta_{AB}^{(N)}(e)$;
\item if $e$ is a network arc with head $v$, then
$\Psi^{(N)}=(\theta^{(N)}(e),RS(v))$.
\end{itemize}
We shall call $\Psi^{(N)}(e)$ the \emph{enriched ($AB$-weighted) tripartition} of $N$ associated to $e$.
Notice that $\Psi^{(N)}(e)$ still depends always only on $e$'s head.

Let us denote by $\Psi(N)$ the set of all these enriched tripartitions:
$$
\Psi(N)=\{\Psi^{(N)}(e)\mid e\in E\}.
$$
Then, in the papers quoted above, these sets $\Psi(N)$ are used to
define a metric $m_{\Psi}$ by means of a formula similar to that of
$m_\Upsilon$ recalled in Section~\ref{sec:trip}.

It is claimed in \cite[\S 5.4]{nakhleh.ea:tutpsb04} and \cite[\S
5]{nakhleh.ea:TR0406} that $\Psi$ satisfies the separation property on
the class of all phylogenetic networks, in the sense that if
$\Psi(N_1)=\Psi(N_2)$ then $N_1\cong N_2$, for every phylogenetic
networks $N_1$ and $N_2$ labeled in the same set $S$.  And again, it
is untrue, as there exist pairs of non-isomorphic phylogenetic
networks with the same sets of enriched tripartitions. For instance,
the phylogenetic networks $N_{3}$ and $N_{4}$  labeled in
$\{1,\ldots,13\}$, used already in the previous section and depicted in
Fig.~\ref{fig:contr2}.  They have the same hybrid
nodes, and we have already seen in Table \ref{tbl:contr2} that they
have the same sets of tripartitions induced by tree arcs as well as
the same sets of tripartitions induced by network arcs.  Table
\ref{tbl:contr3} shows that their hybrid nodes have the same
reticulation scenarios.  From these two tables, one can easily check
that $\Psi(N_{3})=\Psi(N_{4})$.

\section{Nakhleh's thesis: introducing the tree-sibling condition}
\label{sec:phd}

In his PhD Thesis \cite{nakhleh:phd04}, Nakhleh uses the enriched
tripartitions $\Psi$ (actually, he uses unweighted tripartitions, but
for the sake of generality we shall still use them $AB$-weighted), but
he restricts the subclass of phylogenetic networks where $\Psi$ is
stated to satisfy the separation property.

The definition of \emph{model} phylogenetic network in this work is
that of phylogenetic network given in the last section
(conditions (6.1), (6.2) and (6.3)), while in \emph{reconstructible}
networks these conditions are relaxed as in Section \ref{sec:second}.
  
Then he defines a phylogenetic network to be of \emph{class I} when
every hybrid node has at least one sibling that is a tree node.  We
shall say henceforth that such a phylogenetic network satisfies the
\emph{tree-sibling condition}, or simply that it is
\emph{tree-sibling}, to distinguish these networks from previous class
I networks defined through the absence of convergent pairs: notice
that, for instance, phylogenetic networks $N_3$ and $N_4$ in
Fig.~\ref{fig:contr2} have no pair of convergent nodes, but they are
not tree-sibling, while networks $N_5$ and $N_6$ in
Fig.~\ref{fig:contr3}  in the Appendix are tree-sibling, but have pairs of
convergent nodes (see Table \ref{tbl:contr4}).  The phylogenetic
networks used in \cite{nakhleh.ea:bioinfo06,nakhleh.ea:bioinfo07}
(obtained by adding network arcs to a phylogenetic tree by repeating
the following procedure: choose pairs of arcs $(u_1,v_1)$ and
$(u_2,v_2)$ in the tree; split the first into $(u_1,w_1)$ and
$(w_1,v_1)$, with $w_1$ a new (tree) node; split the second one into
$(u_2,w_2)$ and $(w_2,v_2)$, with $w_2$ a new (hybrid) node; finally,
add a new arc $(w_1,w_2)$) are tree-sibling.

Nakhleh claims \cite[Thm.~4 in Ch.~6]{nakhleh:phd04} that $\Psi$
satisfies the separation property on the subclass of all tree-sibling
phylogenetic networks.  It is false, as there exist pairs of
non-isomorphic tree-sibling model phylogenetic networks with the same
sets of enriched tripartitions Consider for instance the phylogenetic
networks $N_{5}$ and $N_{6}$ labeled in $\{1,\ldots,10\}$ depicted in
Fig.~\ref{fig:contr3} in the Appendix.  They are model phylogenetic
networks (they even satisfy the strong time consistency condition
(5.3) instead of the time consistency condition (6.3)), they satisfy
the tree-sibling condition, and they are clearly non-isomorphic.

Table \ref{tbl:contr4} displays the tripartitions of these networks
induced by their arcs, and Table \ref{tbl:contr5} gives the
reticulation scenarios of their hybrid nodes (which are the same in
both networks).  From these two tables, one can easily check that
$\Psi(N_{5})=\Psi(N_{6})$.

\section{Tripartitions  do not distinguish distinguishable networks}
\label{sec:ieee}

In the final paper \cite{moret.ea:2004} of this series, Moret,
Nakhleh, Warnow et al assert that their tripartitions can be used to
distinguish networks up to a certain notion of reduction that we
recall below, from where they deduce that $\theta$ satisfies the
separation property on a very restricted subclass of phylogenetic
networks.

The notion of \emph{model phylogenetic network} is this paper is
exactly that of the last two sections.  As far as {reconstructible}
phylogenetic networks goes, they do not relax them as much as in
previous papers.  More specifically, a \emph{reconstructible
phylogenetic network} on a set $S$ of labels is defined as a rooted
DAG labeled in $S$ satisfying the following conditions:
\begin{enumerate}[(\thesection.1)]
\item The root and all internal tree nodes can have any out-degree
greater than 1.  All hybrid nodes have out-degree 1, but they can have
any in-degree greater than 1.
    
\item The child of a hybrid node is always a tree node.

\item  Time consistency property (6.3).
\end{enumerate}

A subset $U$ of internal nodes of $V$ is said to be \emph{convergent}
when it has at least two elements, and all nodes in it have exactly
the same cluster (contrary to previous versions, no condition on the
number of hybrid nodes in the paths to descendant leaves is required).
The removal of convergent sets is the basis of the following
\emph{reduction procedure}:
\begin{enumerate}
\item[(0)] Replace every clade by a new ``symbolic leaf'' labeled with the
names of all leaves in it.

\item[(1)] For every maximal convergent set $U$, remove all nodes in
paths from nodes in $U$ to (symbolic) leaves, including the
node in $U$ but keeping the leaf.  For every node $x$ that is the tail
of an arc whose head $v$ has been removed, and for every leaf $s$ in
the cluster of $v$, add a new arc $(x,s)$.

(The resulting network contains no convergent set of nodes, because
this step does not change the clusters of the surviving nodes.)

\item[(2)] Append to every symbolic leaf representing
a clade  the corresponding clade, with an arc from the symbolic
leaf to the root of the clade.

\item[(3)] Replace every path of length greater than 1 with all its
intermediate nodes of in- and out-degree equal to 1 by a single arc
from its origin to its end.

(In particular, if a symbolic leaf turns out to have only one parent,
then it is removed and the root of the corresponding clade is appended
to its first ancestor with out-degree different from 1.)
\end{enumerate}

The output of this procedure applied to a reconstructible phylogenetic
network $N$ is a DAG $R(N)$ labeled in $S$.  The network $R(N)$ is
called the \emph{reduced version} of $N$.  Two networks $N_{1}$ and
$N_{2}$ are said to be \emph{indistinguishable} when they have
isomorphic reduced versions, that is, when $R(N_{1})\cong R(N_{2})$.

Since every hybrid node in $N$ and its only child form a convergent
set, they are removed in step (1) together with all their descendants
until the clades' symbolic leaves.  On the other hand, in $R(N)$ the symbolic
leaves may have more than one parent, and then they are the only
possible hybrid nodes in $R(N)$.  So, in particular, no hybrid node in
$R(N)$ is a descendant of another hybrid node. Moreover,
since all convergent sets and all nodes with in- and out-degree 1 in
$N$ are removed, the only possible convergent sets in $R(N)$ consist
of a hybrid node and its only child (that is, a symbolic leaf with
more than one parent and the root of the corresponding clade).

We want to  point out that the reduced version of a
reconstructible phylogenetic network need not be a reconstructible
phylogenetic network.  Consider for instance the simple network $N$ in
Fig.~\ref{fig:reduced1} below.  The nodes $a,b,r$ form a convergent set, and
therefore they are removed in the reduction process.  Then, the reduced
version of $N$ is a non connected DAG consisting of four arcs with
heads the leaves of $N$. As another example, consider the reconstructible
phylogenetic network $N_{8}$ in Fig.~\ref{fig:contr4} in the
Appendix: its reduced
version, which is shown in Fig.~\ref{fig:contr4.2}, does not satisfy
the time consistency property. This shows that
reduced versions need not be
time consistent. 

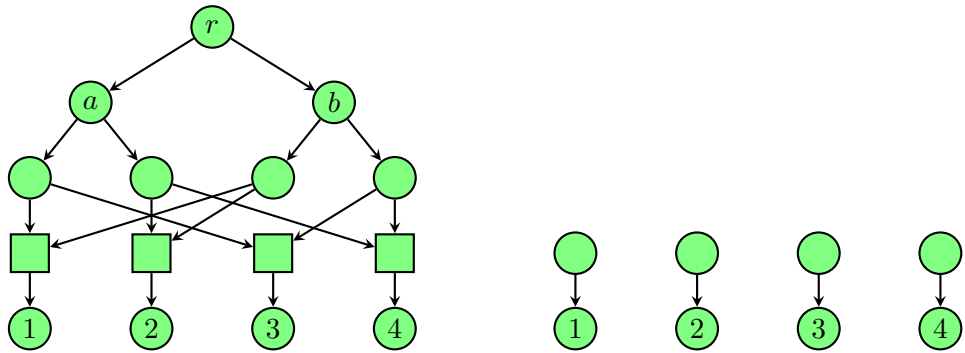
\begin{figure}[htb]
\begin{center}
\begin{tikzpicture}[thick,>=stealth,xscale=0.8]
\draw (3,0) node[tre] (r) {}; \etq r
\draw (1,-1) node[tre] (a) {}; \etq a
\draw (5,-1) node[tre] (b) {}; \etq b
\draw (0,-2) node[tre] (c) {}; 
\draw (2,-2) node[tre] (d) {}; 
\draw (4,-2) node[tre] (e) {}; 
\draw (6,-2) node[tre] (f) {}; 
\draw (0,-3) node[hyb] (C) {}; 
\draw (2,-3) node[hyb] (D) {}; 
\draw (4,-3) node[hyb] (E) {}; 
\draw (6,-3) node[hyb] (F) {}; 
\draw (0,-4) node[tre] (1) {}; \etq 1
\draw (2,-4) node[tre] (2) {}; \etq 2
\draw (4,-4) node[tre] (3) {}; \etq 3
\draw (6,-4) node[tre] (4) {}; \etq 4
\draw [->](r)--(a);
\draw [->](r)--(b);
\draw [->](a)--(c);
\draw [->](a)--(d);
\draw [->](b)--(e);
\draw [->](b)--(f);
\draw [->](c)--(C);
\draw [->](c)--(E);
\draw [->](d)--(D);
\draw [->](d)--(F);
\draw [->](e)--(C);
\draw [->](e)--(D);
\draw [->](f)--(E);
\draw [->](f)--(F);
\draw [->](C)--(1);
\draw [->](D)--(2);
\draw [->](E)--(3);
\draw [->](F)--(4);
\end{tikzpicture}
\qquad\ \qquad
\begin{tikzpicture}[thick,>=stealth,xscale=0.8]
\draw (0,-3) node[tre] (C) {}; 
\draw (2,-3) node[tre] (D) {}; 
\draw (4,-3) node[tre] (E) {}; 
\draw (6,-3) node[tre] (F) {}; 
\draw (0,-4) node[tre] (1) {}; \etq 1
\draw (2,-4) node[tre] (2) {}; \etq 2
\draw (4,-4) node[tre] (3) {}; \etq 3
\draw (6,-4) node[tre] (4) {}; \etq 4
\draw [->](C)--(1);
\draw [->](D)--(2);
\draw [->](E)--(3);
\draw [->](F)--(4);
\end{tikzpicture}
\end{center}
\caption{\label{fig:reduced1} A model phylogenetic network  $N$ (left) 
and its reduced version (right)}
\end{figure}

It is claimed in \cite[Lem.~1]{moret.ea:2004} that if
$\theta(N_1)=\theta(N_2)$, then $N_{1}$ and $N_{2}$ are
indistinguishable (the converse does not hold in general, because the
reduction process may remove parts with different topologies that
yield differences in the sets of tripartitions).  This is not true.
Consider for instance the model phylogenetic networks $N_{7}$ and
$N_{8}$ depicted in Fig.~\ref{fig:contr4} in the Appendix.  Table
\ref{tbl:contr6} displays the tripartitions of these networks induced
by their arcs, showing that the sets of tripartitions are the same (we
give actually the $AB$-weighted tripartitions, just to show that their
claim is still false if we replace $\theta$ by $\theta_{AB}$).
Fig.~\ref{fig:contr4.2} shows the reduced versions of these networks,
and they are clearly non isomorphic.

The authors also claim \cite[Thm.~3]{moret.ea:2004} that $\theta$
satisfies the separation property on the subclass of reduced
reconstructible phylogenetic networks (that is, of reconstructible
phylogenetic networks that remain untouched under the reduction
procedure).  This is also wrong.  Consider for instance the networks
$N_{9}$ and $N_{10}$ depicted in Fig.~\ref{fig:contr6} in the
Appendix.  They are reconstructible phylogenetic networks, and they
are reduced because the only convergent sets they contain consist of a
hybrid node and a leaf that is its only child, and therefore the
application of the reduction procedure leaves them untouched.  Table
\ref{tbl:contr7.2} shows that these two networks have the same sets of
$AB$-weighted tripartitions, but they are clearly non-isomorphic.

\section{Tree-child phylogenetic networks}
\label{sec:treechild}

We have seen in previous sections that neither the lack of convergent
pairs of nodes, nor the tree-sibling condition or even the property of
being reduced, in all cases combined with the strong time consistency
condition, guarantee the separation property for $\theta_{AB}$.  In
this section we introduce a stronger condition that, by itself, does
not guarantee this separation property either, but that combined with
condition (4.3.a) makes $d_{\theta}$  satisfy
the separation property.  Even more, it makes \emph{bipartitions}
$\pi$ satisfy the separation property.

We shall say that a DAG satisfies the \emph{tree-child condition}, or
simply that it is \emph{tree-child}, when every node other than a leaf
has at least one tree child.  A \emph{tree-child phylogenetic network}
is a rooted tree-child DAG with no tree node of out-degree 1 and
all hybrid nodes of out-degree exactly 1 (and any in-degree greater than 1).
So, tree-child phylogenetic networks can be understood as models of
reticulated evolution where:
\begin{itemize}
\item The tree nodes represent species.
\item The hybrid nodes represent recombination or lateral gene
  transfer events that yield the
species corresponding to their single tree child.
\item Every species other that the extant ones, represented by the
leaves, have some descendant through mutation.
\end{itemize}

The (even enriched) tripartitions introduced so far do not satisfy the
separation property on the subclass of \emph{all} tree-child phylogenetic
networks.  Consider for instance the networks $N_{11}$ and $N_{12}$
depicted in Fig.~\ref{fig:treech-nocomp} in the Appendix.
Table~\ref{tbl:treech-nocomp} provides the $AB$-weighted tripartitions
and reticulation scenarios of their arcs, showing that these networks
cannot be distinguished using this information.

But if we add to the tree-child condition the weakest form of time
consistency, namely condition (4.3.a) (two parents of a hybrid node
cannot be connected by a path), then $\theta$ satisfies the separation
property on the resulting subclass of phylogenetic networks.  Actually,
it turns out that the bare sets of clusters of nodes (without
distinguishing between strict and non-strict descendants, and without
taking into account the numbers of hybrid nodes in paths to leaves)
are enough to characterize these phylogenetic networks up to
isomorphism (Thm.~\ref{thm:C}), which entails that $\pi$ satisfies the
separation property, just as in phylogenetic trees.

 To prove these facts, we need to establish
some preliminary definitions and results.

We shall denote henceforth by $C^{(N)}(v)$ the cluster of a node $v$
in a DAG $N$ labeled in $S$, to emphasize the network.  For every DAG
$N=(V,E)$, let $C(N)$ denote the set of clusters of its nodes:
$$
C(N)=\{C^{(N)}(v)\mid v\in V\}.
$$

A \emph{tree path} in a DAG $N$ is a path consisting entirely of tree
arcs.  A node $v$ is a \emph{tree descendant} of a node $u$ when there
exists a tree path from $u$ to $v$.  

\begin{lem}
Every node $v$ in  a tree-child
phylogenetic network  has some tree descendant
leaf.
\end{lem}

\begin{pf}
If $v$ is not already a leaf, we can construct a tree path starting in $v$ by
successively taking tree children, and this path will end in a leaf
that will be a tree descendant of $v$.\qed
\end{pf}

\begin{lem}\label{lem:unicity-path}
Let $u\pathgr v$ be a tree path in a DAG $N$.  Then, for every other
path $w\pathgr v$ ending in $v$, either it contains $u\pathgr v$ or
$u\pathgr v$ contains $w\pathgr v$.
\end{lem}

\begin{pf}
Let $(u,v_{1},\ldots,v_{k-1},v)$ be the tree path $u\pathgr v$ in the
statement: to simplify the notations, we call $v_{0}=u$ and $v_{k}=v$.
Let now $w\pathgr v$ be any other path ending in $v$ and let $v_{j}$ be
the first node in the path $u\pathgr v$ such that $(v_{j},\ldots,v_{k})$
is contained in $w\pathgr v$.  If $j\neq 0$, then $v_{j}$ has only one
parent, and therefore it must happen that either $v_{j}=w$ or
$v_{j}$'s parent in the path $w\pathgr v$ is also its parent in the path
$u\pathgr v$, and in particular it also belongs to this path, which
contradicts the assumption on $v_{j}$.

This proves that either $v_{j}=w$, in which case the path $u\pathgr v$
contains the path $w\pathgr v$, or $v_{j}=u$, in which case the path
$w\pathgr v$ contains the path $u\pathgr v$.\qed
\end{pf}

\begin{cor}\label{cor:unicity-path}
If $v$ is a tree descendant of $u$ in a DAG $N$, then $v\in A(u)$ and
the path $u\pathgr v$ is unique.
\end{cor}

\begin{pf}
If $v$ is a tree descendant of $u$, then there exists a tree path
$u\pathgr v$.  Then, by the previous lemma, every path from a root $r$
to $v$ must contain this path $u\pathgr v$, which shows that $v$ is a
strict descendant of $u$, and every other path from $u$ to $v$ must
contain (and hence be equal to) this path $u\pathgr v$.  \qed
\end{pf}

\begin{lem}\label{lem:C-char}
Let $N=(V,E)$ be a tree-child phylogenetic network satisfying
condition (4.3.a).  For every nodes $u,v\in V$, the following
conditions are equivalent:
\begin{enumerate}[(i)]
\item $C^{(N)}(u)=C^{(N)}(v)$
\item $u=v$ or $\{u,v\}$ are a hybrid node and its only child.
\end{enumerate}
\end{lem}

\begin{pf}
The implication  
(ii)$\Longrightarrow$(i) is obvious.  As far as the 
implication (i)$\Longrightarrow$(ii) goes, assume that $C^{(N)}(u)=C^{(N)}(v)$ and that $u\neq
v$.  If $s$ is a tree descendant leaf of $u$, then $s\in C^{(N)}(v)$,
and hence, by Lemma~\ref{lem:unicity-path}, either $u$ belongs to
the path $v\pathgr s$ or $v$ belongs to the path $u\pathgr s$.
Therefore, $u$ and $v$ are connected by a path.  To fix ideas, assume
that there exists a path $u\pathgr v$.  Then we must distinguish three
cases:
\begin{itemize}
\item If $u$ is a tree node and it has some tree child $w$ outside the
path $u\pathgr v$, then every tree descendant leaf $s$ of $w$ is a tree
descendant leaf of $u$ and hence, since $C^{(N)}(u)=C^{(N)}(v)$, a
descendant of $v$.  By the uniqueness of the path $u\pathgr s$
(Corollary~\ref{cor:unicity-path}), the tree path $u\pathgr s$ must be
equal to the concatenation of the path $u\pathgr v$ and the path $v\pathgr
s$, but these paths are different because their first arcs are
different.  This yields a contradiction.

\item If $u$ is a tree node and all its children outside the path
$u\pathgr v$ are hybrid, take one such hybrid child $w$.  Let $s$ be any
tree descendant leaf of $w$.  Then $s\in C^{(N)}(u)=C^{(N)}(v)$ and
therefore there exists a path $v\pathgr s$.  Now, by
Corollary~\ref{cor:unicity-path}, the path $w\pathgr s$ is unique, and
therefore the path $u\pathgr s$ is also unique.  Indeed, given any path
$u\pathgr s$, by Lemma~\ref{lem:unicity-path} and since $u$ cannot be a
descendant of $w$, the path $w\pathgr s$ must be contained in this path
$u\pathgr s$, and since no other parent of $w$ is a descendant of $u$ by
(4.3.a), the only possibility is that the first arc of this path is
$(u,w)$ and then the rest of the path is the tree path $w\pathgr s$.

This means that the path obtained by concatenating the paths $u\pathgr
v$ and $v\pathgr s$ must be equal to the path $u\pathgr s$ through $w$,
but, again, these paths are different because their first arcs are
different.  This yields again a contradiction.

\item If $u$ is a hybrid node and $u'$ is its unique tree child, then
$C^{(N)}(u')=C^{(N)}(u)=C^{(N)}(v)$ and $u'$ must be the first node
after $u$ in the path $u\pathgr v$.  This yields a path $u'\pathgr v$ with
$C^{(N)}(u')=C^{(N)}(v)$ and $u'$ a tree node.  The last two points
have shown that the assumption that $u'\neq v$ leads to a
contradiction, while $u'=v$ is
clearly possible. 
\end{itemize}
In summary, the only situation that does not lead to a contradiction is when
$u$ is a hybrid node and $v$ its only child.\qed
\end{pf}

\begin{lem}\label{lem:C-ord}
Let $N=(V,E)$ be a tree-child phylogenetic network satisfying
condition (4.3.a).  For every nodes $u,v\in V$ such that $C^{(N)}(u)\neq
C^{(N)}(v)$, the following two conditions are equivalent:
\begin{enumerate}[(i)]
\item There is a non-trivial path $u\pathgr v$.
\item $C^{(N)}(v)\subsetneq C^{(N)}(u)$.
\end{enumerate}
\end{lem}

\begin{pf}
The implication (i)$\Longrightarrow$(ii) is straightforward: if $v$ is
a descendant of $u$, then $C^{(N)}(v)\subseteq C^{(N)}(u)$, and by assumption
these clusters are different.
 
As far as the converse implication goes, assume that
$C^{(N)}(v)\subsetneq C^{(N)}(u)$, and let $s$ be a tree descendant
leaf of $v$.  Then $s\in C^{(N)}(u)$ and Lemma~\ref{lem:unicity-path}
entails that $u$ and $v$ are connected by a path.  Since clusters
decrease with paths, this path must be $u\pathgr v$.  \qed
\end{pf}

Now, given a tree-child phylogenetic network $N=(V,E)$ satisfying
condition (4.3.a), its \emph{contracted
version} is the DAG $N'=(V',E')$ obtained from $N$ by contracting into
one node each pair of nodes consisting of a hybrid node and its only
child: more specifically, for every hybrid node $u$, if $\bar{u}$ is
its only child and $\bar{u}_{1},\ldots,\bar{u}_{k}$ the children of
$\bar{u}$, then we remove this node $\bar{u}$ and the arcs incident to
it, and we replace the latter by new arcs $(u,\bar{u}_{1})$,\ldots,
$(u,\bar{u}_{k})$.  In this way, we understand $V'$ as a subset of
$V$, consisting of all nodes of $N$ except the children of hybrid
nodes.

It is clear that $C^{(N)}(v)=C^{(N')}(v)$ for every node $v\in V'$,
and hence that $C(N)=C(N')$.  Moreover, $C^{(N')}(u)\neq C^{(N')}(v)$
if $u\neq v$, because each pair of nodes in $N$ with the same cluster
has been contracted to a single node in $N'$.  In particular, the
mapping
$$
\begin{array}{rcl}
C^{(N')}: V' & \to & C(N')\\
v & \mapsto & C^{(N')}(v)
\end{array}
$$
is bijective.

On the other hand, for every $u,v\in V'$, there exists a path $u\pathgr
v$ in $N$ if and only if there exists a path $u\pathgr v$ in $N'$.
Therefore, from Lemma \ref{lem:C-ord} we deduce that there exists a
path $u\pathgr v$ in $N'$ if and only if $C^{(N')}(v)\subseteq
C^{(N')}(u)$: that is, the inclusion of clusters in $N'$ captures
exactly the path ordering in $N'$, which is the restriction to $V'$ of
the path ordering on $N$.

\begin{lem}\label{lem:Mu}
Let $N$ be a tree-child phylogenetic network satisfying condition
(4.3.a), and let $N'=(V',E')$ be its contracted version.  For every
$u\in V'$, let
$$
M_{u}=\{w\in V'\mid C^{(N')}(w)\subsetneq C^{(N')}(u)\}.
$$
Then, the maximal elements of $M_{u}$ with respect to the path
ordering on $N'$ are exactly the children of $u$ in $N'$.
\end{lem}

\begin{pf}
If $u$ is a leaf, then $C^{(N')}(u)=\{u\}$ and $M_{u}=\emptyset$, and
the thesis of the statement clearly holds.  So, assume that $u$ is not
a leaf.  Then, every  descendant of $u$ is in $M_{u}$ and therefore $M_{u}$
is non-empty.

Since $M_{u}$ is finite, it has maximal elements.  Let $v$ be any such
a maximal element.  Since $C^{(N')}(v)\subsetneq C^{(N')}(u)$, there
exists a non trivial path $u\pathgr v$ in $N'$.  If this path passes
through some other node $w$, then $C^{(N')}(v)\subsetneq C^{(N')}(w)
\subsetneq C^{(N')}(u)$, against the assumption that $v$ is maximal in
$M_{u}$.  Therefore, the path $u\pathgr v$ has length 1 and $v$ is a
child of $u$. 

Conversely, let $v$ be a child of $u$.  If it is not maximal in
$M_{u}$, then there exists a path $u\pathgr v$ different from the arc
$(u,v)$.  Let $w$ be the parent of $v$ in this path.  Then $v$ has (at
least) two parents, $u$ and $w$, and there is a path $u\pathgr w$ in
$N'$.  When we translate this situation to $N$, we have essentially
four possibilities:
\begin{itemize}
\item $v$ is a hybrid node with parents $u,w$ and there exists a path
$u\pathgr w$ in $N$ connecting them.

\item $v$ is a hybrid node with parents $u$ and the tree child
$\bar{w}$ of the hybrid node $w$.  Then the path $u\pathgr w$ in $N'$
corresponds to a path $u\pathgr w$ in $N$ that, followed by the arc
$(w,\bar{w})$, yields a path $u\pathgr \bar{w}$.

\item $v$ is a hybrid node with parents $w$ and the tree child
$\bar{u}$ of the hybrid node $u$.  Then the path $u\pathgr w$ in $N'$
corresponds to a path $u\pathgr w$ in $N$, and since $\bar{u}$ is the
only child of $u$, the latter contains a path $\bar{u}\to w$.

\item $v$ is a hybrid node with parents the tree child
$\bar{w}$ of the hybrid node $w$ and the tree child
$\bar{u}$ of the hybrid node $u$.  Then the path $u\pathgr w$ in $N'$
corresponds to a path $u\pathgr w$ in $N$. Arguing as in the last two
points (simultaneously), we deduce that there exists a path 
$\bar{u}\pathgr \bar{w}$ in $N$.
\end{itemize}
In all four cases, we obtain a hybrid node of $N$ and a path
connecting two parents of it, which contradicts (4.3.a).
This implies that $v$ must be maximal in $M_{u}$.
\qed
\end{pf}

\begin{thm}\label{thm:C}
For every tree-child phylogenetic networks $N_{1}$ and $N_{2}$
satisfying the weak time consistency property (4.3.a),
$$
N_{1}\cong N_{2} \mbox{ if and only if } C(N_{1})=C(N_{2}).
$$
\end{thm}

\begin{pf}
Assume that $C(N_{1})=C(N_{2})$, and
let $N_{1}'=(V_{1}',E_{1}')$ and $N_{2}'=(V_{2}',E_{2}')$ be the
contracted versions of $N_{1}$ and $N_{2}$. Then $C(N_{1}')=C(N_{2}')$.

Consider  the mapping  $f': V_{1}' \longrightarrow V_{2}'$ obtained 
as the composition
$$
V_{1}' \stackrel{C^{(N'_{1})}}{\longrightarrow} C(N_{1}')=C(N_{2}')
\stackrel{{C^{(N'_{2})}}^{-1}}{\longrightarrow} V_{2}';
$$
that is, $f'$ sends each node $v\in V_{1}'$ to the unique node $f'(v)\in
V_{2}'$ such that $C^{(N'_{1})}(v)=C^{(N'_{2})}(f'(v))$.  Since
$C^{(N'_{1})}$ and $C^{(N'_{2})}$ are bijective, $f'$ is bijective.
Furthermore, $v$ is maximal in $M_{u}$ if and only if $f'(v)$ is
maximal in $M_{f'(u)}$ (because these sets are defined through the
corresponding clusters, and the path ordering in $N_{1}'$ and $N_{2}'$
corresponds to the inclusion of clusters). Therefore,
$(u,v)\in E_{1}'$ if and only if $(f'(u),f'(v))\in E_{2}'$. So, $f$ is
an isomorphism of DAGs.

Finally, $u$ is a leaf of a DAG if and only if its cluster is the
singleton $\{u\}$.  This entails that $f'$ sends leaves to leaves and
preserves their labels.  Therefore, $f'$ is an isomorphism of DAGs
labeled in $S$.

Now, the hybrid nodes in each $N_{i}$ are the nodes that have
in-degree greater than 1 in the corresponding $N_{i}'$, and $N_{i}$ is
obtained from $N_{i}'$ by adding a single child $\bar{u}$ to each
hybrid node $u$ and replacing all arcs with tail $u$ by arcs with tail
$\bar{u}$ (and the same heads). This implies that the mapping
$$
f:V_{1}\to V_{2}
$$
that restricts to $f'$ on $V_{1}'$ and that sends each node $\bar{u}$
in $V_{1}\setminus V'_{1}$ to the corresponding $\overline{f'(u)}$ (that is, 
to the only child of the image of the parent of $\bar{u}$)  is
bijective and preserves and reflects the arcs and preserves the
leaves' labels. Therefore, it is an isomorphism of phylogenetic
networks.
\qed
\end{pf}

\begin{cor}\label{prop:tc-error}
The bipartition $\pi$, and hence also $\theta$, $\theta_{B}$,
$\theta_{AB}$, and $\Psi$, satisfy the separation property on the
subclass of all tree-child phylogenetic networks where no pair of parents
of a hybrid node is connected by a path.
\end{cor}

\begin{cor}\label{prop:RF-netw}
For every $\Upsilon=\pi,\theta,\theta_{B},\theta_{AB},\Psi$, the mapping
$$
d_{\Upsilon}(N_{1},N_{2})
=\frac{1}{2}\big(|\Upsilon(T)\setminus \Upsilon(T')|+|\Upsilon(T')\setminus
\Upsilon(T)|\big)
$$
defines a distance on the
subclass of all tree-child phylogenetic networks where no pair of parents
of a hybrid node is connected by a path.
\end{cor}

\section{Conclusion}

In a series of technical reports and papers culminating in
\cite{moret.ea:2004}, Moret, Nakhleh, Warnow and collaborators have
introduced an error metric for phylogenetic networks, with the main
goal of comparing reconstructed networks with true ones and to assess
in this way the accuracy of phylogenetic network reconstruction
algorithms.  In this paper we have shown that none of their approaches
is free from false equalities: that is, for every one of the metrics
they introduce, there turn out to exist pairs of phylogenetic networks
in the subclass where the metric is defined, that are non-isomorphic
(or, in the case of \cite{moret.ea:2004}, have non-isomorphic reduced
versions) but cannot be distinguished through the metric.  The reason
for this lack of discriminating power is that non-isomorphic networks
in the subclasses under consideration may have the same sets of
tripartitions, which are the networks' representations that are
compared by this metric.  Among these subclasses of networks where the
error metric fails, we want to stress the \emph{tree-sibling,
strongly time consistent phylogenetic networks} (see
Section~\ref{sec:phd}), for which several reconstruction algorithms
were recently proposed
\cite{nakhleh.ea:bioinfo06,nakhleh.ea:bioinfo07}.

We have also shown a subclass of phylogenetic networks, the
\emph{tree-child, weakly time consistent phylogenetic networks}, where
tripartitions, and even bipartitions in the sense of
Bourque-Robinson-Foulds, single out its members, and therefore they can
be used to define a true metric. Tree-child phylogenetic networks can
be seen as models of reticulate evolution histories where every
species other than the extant ones have some descendant through
mutation. 

Several questions and problems arise as a consequence of our work that
are in our current research agenda.  On the one hand, what is the
discriminating power of tripartitions?  Is there a well-defined class
of phylogenetic networks where equality of sets, or multisets (sets
with repetitions), of tripartitions implies isomorphism?  And, what
does it really mean, from the topological point of view, to have the
same (multi)sets of tripartitions?

On the other hand, it is still necessary to define true
distances generalizing the bipartition distance, on more general
subclasses of phylogenetic networks than those tree-child, weakly time
consistent.  We have recently defined one such metric (not
based on tripartitions) on the class of \emph{all} tree-child
phylogenetic networks \cite{cardona.ea:07b}, but, in the light of
\cite{nakhleh.ea:bioinfo06,nakhleh.ea:bioinfo07},
we consider a more
relevant target the class of all tree-sibling networks.

\section*{Appendix}

In this Appendix we collect all depictions of phylogenetic networks
and their tripartitions. 

In graphical representations of phylogenetic networks, and of DAGs in general,
hybrid nodes are represented by squares and tree nodes by circles.
In those cases where the strong time consistency condition is
considered, nodes are labelled with its corresponding $\tau$ (see
Proposition~\ref{prop:timing}) as
subscript, to ease the verification of this condition.

In the tables
presenting tripartitions we shall make use for simplicity of the
following conventions: we only provide the sets $A$ and $B$, as $C^c$
can be trivially deduced from them; the labels' weights are shown as
subscripts, the lack of subscript meaning weight 0; and since the
tripartition induced by an arc only depends on its head, we identify
the arcs by means of their heads.

\begin{table}[htb]
{\scriptsize  \begin{tabular}{|c|c|c|c|c|}
     \hline
        \multicolumn{1}{|c}{arc's}& \multicolumn{2}{|c}{$N_{1}$} &  \multicolumn{2}{|c|}{$N_{2}$}  \\
  \cline{2-5}
      head &\quad $A$\quad{} & $B$ &\quad $A$\quad{} & $B$   \\
      \hline
      $a$ & 1 & $2_{3},3_{2},4_{1}$ & 1 & $2_{3},3_{2},4_{1}$ \\ 
      \hline
      $b$ & 1 & $2_{3},3_{2},4_{1}$  & 1 & $2_{3},3_{2},4_{1}$ \\ 
      \hline 
      $c$ & 1 & $2_{3},3_{2},4_{1}$  & 1 & $2_{3},3_{2},4_{1}$ \\ 
      \hline  
      $d$ & 5 & $2_{3},3_{2},4_{1}$ & 5 & $2_{3},3_{2},4_{1}$ \\ 
      \hline 
      $e$ & 5 & $2_{3},3_{2},4_{1}$ & 5 & $2_{3},3_{2},4_{1}$ \\ 
      \hline 
      $f$ & 5 & $2_{3},3_{2},4_{1}$ & 5 & $2_{3},3_{2},4_{1}$ \\ 
      \hline 
        $g$ & 4 & $2_{2},3_{1}$ &  4 & $2_{2},3_{1}$ \\
       \hline
             $h$ & 3 & $2_{1}$ & 3 & $2_{1}$\\
\hline
        $r$ & $1,2_{3},3_{2},4_{1},5$ & $\emptyset$ & $1,2_{3},3_{2},4_{1},5$ & $\emptyset$ \\
      \hline
    $A$ & $\emptyset$ & $2_{2}$ & $\emptyset$ & $2_{3},3_{2}$ \\
      \hline  
      $B$ & $\emptyset$ & $2_{3},3_{2}$ & $\emptyset$ & $2_{2}$  \\
      \hline
      $C$ & $4_{1}$ & $2_{3},3_{2}$ & $4_{1}$ & $2_{3},3_{2}$  \\
      \hline
      $D$ & $3_{1}$ & $2_{2}$ & $3_{1}$ & $2_{2}$  \\
      \hline
  $E$ & $2_{1}$ & $\emptyset$ & $2_{1}$ & $\emptyset$ \\
 \hline
 \end{tabular} 
}\smallskip
    \caption{Tripartitions of the networks in Fig.~\ref{fig:contr1}}
    \label{tbl:contr1}
\end{table}

\begin{figure}[htb]
\begin{center}
\begin{tikzpicture}[thick,>=stealth,xscale=0.75,yscale=0.85]
\draw (3,\nvl) node[tre] (3) {}; \etq 3 {}
\draw (4,\nvl) node[tre] (4) {}; \etq 4 {}
\draw (5,\nvl) node[tre] (5) {}; \etq 5 {}
\draw (6,\nvl) node[tre] (6) {}; \etq 6 {}
\draw (7,\nvl) node[tre] (7) {}; \etq 7 {}
\draw (8,\nvl) node[tre] (8) {}; \etq 8 {}
\draw (9,\nvl) node[tre] (9) {}; \etq 9 {}
\draw (10,\nvl) node[tre] (10) {}; \etq {10} {}
\draw (11,\nvl) node[tre] (11) {}; \etq {11} {}
\nounivell
\draw (4,\nvl) node[hyb] (J) {}; \etqb J 8
\draw (8,\nvl) node[tre] (v) {}; \etqb v 8
\nounivell
\draw (8,\nvl) node[hyb] (I) {}; \etqb I 7
\draw (10,\nvl) node[tre] (u) {}; \etqb u 7
\nounivell
\draw (5,\nvl) node[hyb] (G) {}; \etqb G 8
\draw (9,\nvl) node[hyb] (H) {}; \etqb H 7
\nounivell
\draw (6,\nvl) node[tre] (s) {}; \etqb s 8
\draw (8,\nvl) node[tre] (t) {}; \etqb t 7
\nounivell 
\draw (4,\nvl) node[tre] (o) {}; \etqb o 8
\draw (7,\nvl) node[tre] (p) {}; \etqb p 6
\draw (9,\nvl) node[tre] (q) {}; \etqb q 7
\nounivell
\draw (3,\nvl) node[tre] (a) {}; \etqb a 5
\draw (10,\nvl) node[hyb] (F) {}; \etqb F 5
\draw (11,\nvl) node[tre] (l) {}; \etqb l 5
\nounivell
\draw (4,\nvl) node[tre] (m) {}; \etqb m 5
\draw (7,\nvl) node[hyb] (E) {}; \etqb E 5
\draw (9,\nvl) node[tre] (n) {}; \etqb n 5
\nounivell
\draw (3,\nvl) node[tre] (b) {}; \etqb b 4
\draw (10,\nvl) node[hyb] (D) {}; \etqb D 4
\draw (11,\nvl) node[tre] (k) {}; \etqb k 4
\nounivell
\draw (1,\nvl)  node[tre] (1) {}; \etq 1 {}
\draw (2,\nvl)  node[tre] (2) {}; \etq 2 {}
\draw (12,\nvl)  node[tre] (12) {}; \etq {12} {}
\draw (13,\nvl)  node[tre] (13) {}; \etq {13} {}
\draw (3,\nvl) node[hyb] (A) {}; \etqb A 3
\draw (11,\nvl) node[hyb] (B) {}; \etqb B 3
\nounivell 
\draw (3,\nvl) node[tre] (c) {}; \etqb c 3
\draw (4,\nvl) node[hyb] (C) {}; \etqb C 3
\draw (11,\nvl) node[tre] (j) {}; \etqb j 3
\nounivell
\draw (1,\nvl) node[tre] (e) {}; \etqb e 3
\draw (3,\nvl) node[tre] (d) {}; \etqb d 2
\draw (11,\nvl) node[tre] (i) {}; \etqb i 2
\draw (13,\nvl) node[tre] (h) {}; \etqb h 3
\nounivell
\draw (5,\nvl) node[tre] (f) {}; \etqb f 1
\draw (9,\nvl) node[tre] (g) {}; \etqb g 1
\nounivell
\draw (7,\nvl)+(0,-0.5) node[tre] (r) {}; \etqb r 0
\draw [->](r)--(f);
\draw [->](r)--(g);
\draw [->](f)--(e);
\draw [->](f)--(d);
\draw [->](g)--(i);
\draw [->](g)--(h);
\draw [->](e)--(1);
\draw [->](e)--(A);
\draw [->](d)--(2);
\draw [->](d)--(c);
\draw [->](i)--(j);
\draw [->](i)--(12);
\draw [->](h)--(B);
\draw [->](h)--(13);
\draw [->](c)--(A);
\draw [->](c)--(C);
\draw [->](C)--(m);
\draw [->](j)--(C);
\draw [->](j)--(B);
\draw [->](A)--(b);
\draw [->](B)--(k);
\draw [->](b)--(a);
\draw [->](b)--(D);
\draw [->](D)--(n);
\draw [->](k)--(D);
\draw [->](k)--(l);
\draw [->](m)--(o);
\draw [->](m)--(E);
\draw [->](E)--(p);
\draw [->](n)--(E);
\draw [->](n)--(q);
\draw [->](a)--(F);
\draw [->](a)--(3);
\draw [->](F)--(u);
\draw [->](l)--(F);
\draw [->](l)--(11);
\draw [->](o)--(J);
\draw [->](o)--(G);
\draw [->](p)--(s);
\draw [->](p)--(t);
\draw [->](q)--(H);
\draw [->](q)--(I);
\draw [->](s)--(G);
\draw [->](s)--(6);
\draw [->](t)--(7);
\draw [->](t)--(H);
\draw [->](G)--(5);
\draw [->](H)--(9);
\draw [->](u)--(I);
\draw [->](u)--(10);
\draw [->](I)--(v);
\draw [->](v)--(J);
\draw [->](v)--(8);
\draw [->](J)--(4);
\end{tikzpicture}
\quad
\begin{tikzpicture}[thick,>=stealth,xscale=0.75,yscale=0.85]
\draw (3,\nvl) node[tre] (3) {}; \etq 3
\draw (4,\nvl) node[tre] (4) {}; \etq 4
\draw (9,\nvl) node[red,tre] (5) {}; \draw[red] (5) node {$5$};
\draw (7,\nvl) node[red,tre] (6) {}; \draw[red] (6) node {$6$};
\draw (6,\nvl) node[red,tre] (7) {}; \draw[red] (7) node {$7$};
\draw (8,\nvl) node[tre] (8) {}; \etq 8
\draw (5,\nvl) node[red,tre] (9) {}; \draw[red] (9) node {$9$};
\draw (10,\nvl) node[tre] (10) {}; \etq {10}
\draw (11,\nvl) node[tre] (11) {}; \etq {11}
\nounivell
\draw (4,\nvl) node[hyb] (J) {}; \etqb J 8
\draw (8,\nvl) node[tre] (v) {}; \etqb v 8
\nounivell
\draw (8,\nvl) node[hyb] (I) {}; \etqb I 7
\draw (10,\nvl) node[tre] (u) {}; \etqb u 7
\nounivell
\draw (5,\nvl) node[hyb] (G) {}; \etqb G 7
\draw (9,\nvl) node[hyb] (H) {}; \etqb H 8
\nounivell
\draw (6,\nvl) node[tre] (s) {}; \etqb s 7
\draw (8,\nvl) node[tre] (t) {}; \etqb t 8
\nounivell 
\draw (4,\nvl) node[tre] (o) {}; \etqb o 7
\draw (7,\nvl) node[tre] (p) {}; \etqb p 6
\draw (9,\nvl) node[tre] (q) {}; \etqb q 8
\nounivell
\draw (3,\nvl) node[tre] (a) {}; \etqb a 5
\draw (10,\nvl) node[hyb] (F) {}; \etqb F 5
\draw (11,\nvl) node[tre] (l) {}; \etqb l 5
\nounivell
\draw (4,\nvl) node[tre] (m) {}; \etqb m 5
\draw (7,\nvl) node[hyb] (E) {}; \etqb E 5
\draw (9,\nvl) node[tre] (n) {}; \etqb n 5
\nounivell
\draw (3,\nvl) node[tre] (b) {}; \etqb b 4
\draw (10,\nvl) node[hyb] (D) {}; \etqb D 4
\draw (11,\nvl) node[tre] (k) {}; \etqb k 4
\nounivell
\draw (1,\nvl)  node[tre] (1) {}; \etq 1
\draw (2,\nvl)  node[tre] (2) {}; \etq 2
\draw (12,\nvl)  node[tre] (12) {}; \etq {12} 
\draw (13,\nvl)  node[tre] (13) {}; \etq {13}
\draw (3,\nvl) node[hyb] (A) {}; \etqb A 3
\draw (11,\nvl) node[hyb] (B) {}; \etqb B 3
\nounivell 
\draw (3,\nvl) node[tre] (c) {}; \etqb c 3
\draw (4,\nvl) node[hyb] (C) {}; \etqb C 3
\draw (11,\nvl) node[tre] (j) {}; \etqb j 3
\nounivell
\draw (1,\nvl) node[tre] (e) {}; \etqb e 3
\draw (3,\nvl) node[tre] (d) {}; \etqb d 2
\draw (11,\nvl) node[tre] (i) {}; \etqb i 2
\draw (13,\nvl) node[tre] (h) {}; \etqb h 3
\nounivell
\draw (5,\nvl) node[tre] (f) {}; \etqb f 1
\draw (9,\nvl) node[tre] (g) {}; \etqb g 1
\nounivell
\draw (7,\nvl)+(0,-0.5) node[tre] (r) {}; \etqb r 0
\draw [->](r)--(f);
\draw [->](r)--(g);
\draw [->](f)--(e);
\draw [->](f)--(d);
\draw [->](g)--(i);
\draw [->](g)--(h);
\draw [->](e)--(1);
\draw [->](e)--(A);
\draw [->](d)--(2);
\draw [->](d)--(c);
\draw [->](i)--(j);
\draw [->](i)--(12);
\draw [->](h)--(B);
\draw [->](h)--(13);
\draw [->](c)--(A);
\draw [->](c)--(C);
\draw [->](C)--(m);
\draw [->](j)--(C);
\draw [->](j)--(B);
\draw [->](A)--(b);
\draw [->](B)--(k);
\draw [->](b)--(a);
\draw [->](b)--(D);
\draw [->](D)--(n);
\draw [->](k)--(D);
\draw [->](k)--(l);
\draw [->](m)--(o);
\draw [->](m)--(E);
\draw [->](E)--(p);
\draw [->](n)--(E);
\draw [->](n)--(q);
\draw [->](a)--(F);
\draw [->](a)--(3);
\draw [->](F)--(u);
\draw [->](l)--(F);
\draw [->](l)--(11);
\draw [red,->](o)--(I);
\draw [->](o)--(G);
\draw [->](p)--(s);
\draw [->](p)--(t);
\draw [->](q)--(H);
\draw [red,->](q)--(J);
\draw [->](s)--(G);
\draw [->](s)--(7);
\draw [->](t)--(6);
\draw [->](t)--(H);
\draw [->](G)--(9);
\draw [->](H)--(5);
\draw [->](u)--(I);
\draw [->](u)--(10);
\draw [->](I)--(v);
\draw [->](v)--(J);
\draw [->](v)--(8);
\draw [->](J)--(4);
\end{tikzpicture}
\end{center}
\caption{\label{fig:contr2} The networks $N_{3}$ (up) and $N_{4}$
(down)}
\end{figure}

 \begin{table}[htb]
{\scriptsize  \begin{tabular}{|c|c|c|c|c|}
      \hline
        \multicolumn{1}{|c}{arc's}& \multicolumn{2}{|c}{$N_{3}$} & 
        \multicolumn{2}{|c|}{$N_{4}$}  \\
  \cline{2-5}
      head &\quad $A$\quad{} & $B$ &\quad $A$\quad{} & $B$   \\
       \hline
$a$ & 3 & $4_{3},8_{2},10_{1}$ &  3 & $4_{3},8_{2},10_{1}$ \\  
\hline
$b$ & 3 & $4_{3},5_{3},6_{2},7_{2},8_{2},9_{3},10_{1}$ & 3 &  $4_{3},5_{3},6_{2},7_{2},8_{2},9_{3},10_{1}$ \\  
\hline
$c$ & $\emptyset$ & $3_{1},4_{4},5_{4},6_{3},7_{3},8_{3},9_{4},10_{2}$ & 
$\emptyset$ & $3_{1},4_{4},5_{4},6_{3},7_{3},8_{3},9_{4},10_{2}$ \\ 
\hline
$d$ & 2 & $3_{1},4_{4},5_{4},6_{3},7_{3},8_{3},9_{4},10_{2}$ & 
2 & $3_{1},4_{4},5_{4},6_{3},7_{3},8_{3},9_{4},10_{2}$ \\  
\hline
$e$ & 1 & $3_{1},4_{4},5_{4},6_{3},7_{3},8_{3},9_{4},10_{2}$ & 1 &
$3_{1},4_{4},5_{4},6_{3},7_{3},8_{3},9_{4},10_{2}$ \\  
\hline
$f$ & $1,2,3_{1}$ & $4_{4},5_{4},6_{3},7_{3},8_{3},9_{4},10_{2}$ & 
$1,2,3_{1}$ & $4_{4},5_{4},6_{3},7_{3},8_{3},9_{4},10_{2}$ \\  
\hline
$g$ & $11_{1},12,13$ & $4_{4},5_{4},6_{3},7_{3},8_{3},9_{4},10_{2}$ & 
$11_{1},12,13$ & $4_{4},5_{4},6_{3},7_{3},8_{3},9_{4},10_{2}$ \\
\hline
$h$ & $13$ & $4_{4},5_{4},6_{3},7_{3},8_{3},9_{4},10_{2},11_{1}$ & 
$13$ & $4_{4},5_{4},6_{3},7_{3},8_{3},9_{4},10_{2},11_{1}$ \\  
\hline
$i$ & $12$ & $4_{4},5_{4},6_{3},7_{3},8_{3},9_{4},10_{2},11_{1}$ & $12$ & $4_{4},5_{4},6_{3},7_{3},8_{3},9_{4},10_{2},11_{1}$ \\  
\hline
$j$ & $\emptyset$ & $4_{4},5_{4},6_{3},7_{3},8_{3},9_{4},10_{2},11_{1}$ & 
$\emptyset$ & $4_{4},5_{4},6_{3},7_{3},8_{3},9_{4},10_{2},11_{1}$ \\  
\hline
$k$ & 11 & $4_{3},5_{3},6_{2},7_{2},8_{2},9_{3},10_{1}$ & 11 & $4_{3},5_{3},6_{2},7_{2},8_{2},9_{3},10_{1}$ \\  
\hline
$l$ & 11 & $4_{3},8_{2},10_{1}$ & 11 & $4_{3},8_{2},10_{1}$ \\  
\hline
$m$ & $\emptyset$ & $4_{1},5_{2},6_{1},7_{1},9_{2}$ & $\emptyset$ & $4_{2},5_{2},6_{1},7_{1},8_{1},9_{2}$ \\
\hline
$n$ & $\emptyset$ & $4_{2},5_{2},6_{1},7_{1},8_{1},9_{2}$ & $\emptyset$ & $4_{1},5_{2},6_{1},7_{1},9_{2}$\\  
\hline
$o$ & $\emptyset$ & $4_{1},5_{1}$ & $\emptyset$ & $4_{2},8_{1},9_{1}$ \\  
\hline
$p$ & $6,7$ & $5_{1},9_{1}$ & $6,7$ & $5_{1},9_{1}$ \\  
\hline
$q$ & $\emptyset$ & $4_{2},8_{1},9_{1}$ & $\emptyset$ & $4_{1},5_{1}$\\  
\hline
$s$ & 6 & $5_{1}$ & 7 & $9_{1}$ \\  
\hline
$t$  & 7 & $9_{1}$ & 6 & $5_{1}$ \\  
\hline
$u$ & 10 & $4_{2},8_{1}$ & 10 & $4_{2},8_{1}$\\  
\hline
$v$ & 8 & $4_{1}$ & 8 & $4_{1}$\\  
\hline
& $1,2, 3_{1},4_{4},5_{4},$ & & 
$1,2, 3_{1},4_{4},5_{4},$ & \\
$r$  & $6_{3},7_{3},8_{3},9_{4},$ & $\emptyset$ & $6_{3},7_{3},8_{3},9_{4},$ &$\emptyset$
  \\
& $10_{2},11_{1},12,13$ & &  $10_{2},11_{1},12,13$ & \\
\hline
$A$ & $3_{1}$ & $4_{4},5_{4},6_{3},7_{3},8_{3},9_{4},10_{2}$  &  $3_{1}$ & $4_{4},5_{4},6_{3},7_{3},8_{3},9_{4},10_{2}$ \\
\hline
$B$ & $11_{1}$ & $4_{4},5_{4},6_{3},7_{3},8_{3},9_{4},10_{2}$ & $11_{1}$ & $4_{4},5_{4},6_{3},7_{3},8_{3},9_{4},10_{2}$ \\
\hline
$C$ & $\emptyset$ & $4_{2},5_{3},6_{2},7_{2},9_{3}$ & $\emptyset$ &
$4_{3},5_{3},6_{2},7_{2},8_{2},9_{3}$  \\
\hline
$D$ & $\emptyset$ &$4_{3},5_{3},6_{2},7_{2},8_{2},9_{3}$ & $\emptyset$ & $4_{2},5_{3},6_{2},7_{2},9_{3}$ \\
\hline
$E$ & $6_{1},7_{1}$ & $5_{2},9_{2}$ & $6_{1},7_{1}$ & $5_{2},9_{2}$ \\
\hline
$F$ & $10_{1}$ & $4_{3},8_{2}$ & $10_{1}$ & $4_{3},8_{2}$\\
\hline
$G$ & $5_{1}$ & $\emptyset$ & $9_{1}$ & $\emptyset$\\
\hline
$H$ & $9_{1}$ & $\emptyset$ & $5_{1}$ & $\emptyset$\\
\hline
$I$ & $8_{1}$ & $4_{2}$ & $8_{1}$ & $4_{2}$\\
\hline
$J$ & $4_{1}$ & $\emptyset$ & $4_{1}$ & $\emptyset$\\
\hline
  \end{tabular} 
 }   \smallskip
   \caption{Tripartitions of the networks in Fig.~\ref{fig:contr2}}
 \label{tbl:contr2}
\end{table}

\begin{table}[htb]
{\scriptsize  \begin{tabular}{|c|c|c|}
     \hline
arc's& $N_{3}$ &  $N_{4}$  \\
 \cline{2-3}
     head & $RS$ & $RS$   \\
      \hline
$A$  & \hspace*{-7ex}$\big\{\{1,3,4,5,6,7,8,9,10\},$ &
\hspace*{-7ex}$\big\{\{1,3,4,5,6,7,8,9,10\},$\\
 & $\qquad\qquad \qquad \{3,4,5,6,7,8,9,10\}\big\}$ &
$\qquad\qquad \qquad \{3,4,5,6,7,8,9,10\}\big\}$\\
\hline
$B$  & \hspace*{-7ex}$\big\{\{4,5,6,7,8,9,10,11,13\}$ &
\hspace*{-7ex}$\big\{\{4,5,6,7,8,9,10,11,13\}$ \\
& $\qquad\qquad \qquad\{4,5,6,7,8,9,10,11\}\big\}$ &
$\qquad\qquad \qquad\{4,5,6,7,8,9,10,11\}\big\}$ \\
\hline
$C$   & \hspace*{-7ex}$\big\{\{3,4,5,6,7,8,9,10\}$ & 
\hspace*{-7ex}$\big\{\{3,4,5,6,7,8,9,10\}$  \\
  & $\qquad\qquad \qquad\{4,5,6,7,8,9,10,11\}\big\}$ & 
$\qquad\qquad \qquad\{4,5,6,7,8,9,10,11\}\big\}$  \\
\hline
$D$   & \hspace*{-7ex}$\big\{\{3,4,5,6,7,8,9,10\}$ & 
\hspace*{-7ex}$\big\{\{3,4,5,6,7,8,9,10\}$  \\
  & $\qquad\qquad \qquad\{4,5,6,7,8,9,10,11\}\big\}$ & 
$\qquad\qquad \qquad\{4,5,6,7,8,9,10,11\}\big\}$  \\
\hline
$E$  & $\big\{\{4,5,6,7,9\},\{4,5,6,7,8,9\}\big\}$ &
$\big\{\{4,5,6,7,9\},\{4,5,6,7,8,9\}\big\}$ \\
\hline
$F$  & $\big\{\{3,4,8,10\},\{4,8,10,11\}\big\}$ &
$\big\{\{3,4,8,10\},\{4,8,10,11\}\big\}$ \\
\hline
$G$  & $\big\{\{4,5\},\{5,6\}\big\}$ & $\big\{\{4,8,9\},\{7,9\}\big\}$\\
\hline
$H$  &  $\big\{\{4,8,9\},\{7,9\}\big\}$ & $\big\{\{4,5\},\{5,6\}\big\}$\\
\hline
$I$ & $\big\{\{4,8,9\},\{4,8,10\}\big\}$ & $\big\{\{4,8,9\},\{4,8,10\}\big\}$\\
\hline
$J$ & $\big\{\{4,5\},\{4,8\}\big\}$ & $\big\{\{4,5\},\{4,8\}\big\}$\\
\hline
\end{tabular} 
}      
\smallskip
\caption{Reticulation scenarios of the hybrid nodes of the networks in Fig.~\ref{fig:contr2}}
\label{tbl:contr3}
\end{table}

\begin{figure}[htb]
\begin{center}
\begin{tikzpicture}[thick,>=stealth,xscale=0.6,yscale=0.8]
\draw (4,0) node[tre] (r) {}; \etqb r 0
\draw (2,-1) node[tre] (c) {}; \etqb c 1
\draw (6,-1) node[tre] (n) {}; \etqb n 1
\draw (4,-2) node[hyb] (A) {}; \etqb A 1
\draw (0,-3) node[tre] (b) {}; \etqb b 2
\draw (9,-3) node[tre] (o) {}; \etqb o 2
\draw (1,-4) node[tre] (d) {}; \etqb d 5
\draw (6,-3) node[hyb] (B) {}; \etqb B 2
\draw (2,-5) node[tre] (f) {}; \etqb f 6
\draw (8,-5) node[tre] (e) {}; \etqb e 4
\draw (2,-6) node[tre] (h) {}; \etqb h 7
\draw (3,-6) node[hyb] (C) {}; \etqb C 6
\draw (6,-6) node[tre] (g) {}; \etqb g 6
\draw (3,-8) node[tre] (i) {}; \etqb i 7
\draw (7,-7) node[tre] (j) {}; \etqb j 7
\draw (2,-9.5) node[hyb] (D) {}; \etqb D 7
\draw (4,-7) node[hyb] (E) {}; \etqb E 7
\draw (4,-9) node[tre] (k) {}; \etqb k 8
\draw (6,-9) node[tre] (l) {}; \etqb l 8
\draw (7,-9) node[tre] (m) {}; \etqb m 8
\draw (0,-10) node[tre] (a) {}; \etqb a 3
\draw (5,-10) node[hyb] (F) {}; \etqb F 8
\draw (7,-10) node[hyb] (G) {}; \etqb G 8
\draw (9,-10) node[tre] (p) {}; \etqb p 3
\draw (8,-11) node[hyb] (H) {}; \etqb H 3
\draw (8,-12) node[tre] (q) {}; \etqb q 4
\draw (3,-12) node[hyb] (I) {}; \etqb I 4
\draw (3,-13) node[tre] (s) {}; \etqb s 5
\draw (1,-13) node[hyb] (J) {}; \etqb J 5
\draw (0,-14) node[tre] (1) {}; \etq 1
\draw (1,-14) node[tre] (2) {}; \etq 2 
\draw (2,-14) node[tre] (3) {}; \etq 3
\draw (3,-14) node[tre] (4) {}; \etq 4
\draw (4,-14) node[tre] (5) {}; \etq 5
\draw (5,-14) node[tre] (6) {}; \etq 6
\draw (6,-14) node[tre] (7) {}; \etq 7
\draw (7,-14) node[tre] (8) {}; \etq 8
\draw (8,-14) node[tre] (9) {}; \etq 9
\draw (9,-14) node[tre] (10) {}; \etq {10}
\draw [->](r)--(c);
\draw [->](r)--(n);
\draw [->](c)--(b);
\draw [->](c)--(A);
\draw [->](n)--(A);
\draw [->](n)--(o);
\draw [->](A)--(d);
\draw [->](b)--(a);
\draw [->](b)--(B);
\draw [->](o)--(B);
\draw [->](o)--(p);
\draw [->](d)--(f);
\draw [->](d)--(J);
\draw [->](B)--(e);
\draw [->](f)--(h);
\draw [->](f)--(C);
\draw [->](e)--(g);
\draw [->](e)--(I);
\draw [->](h)--(D);
\draw [->](h)--(E);
\draw [->](C)--(i);
\draw [->](g)--(C);
\draw [->](g)--(j);
\draw [->](i)--(D);
\draw [->](i)--(k);
\draw [->](j)--(E);
\draw [->](j)--(m);
\draw [->](D)--(3);
\draw [->](E)--(l);
\draw [->](k)--(5);
\draw [->](k)--(F);
\draw [->](l)--(7);
\draw [->](l)--(G);
\draw [->](m)--(F);
\draw [->](m)--(G);
\draw [->](F)--(6);
\draw [->](G)--(8);
\draw [->](a)--(1);
\draw [->](a)--(H);
\draw [->](p)--(H);
\draw [->](p)--(10);
\draw [->](H)--(q);
\draw [->](q)--(I);
\draw [->](q)--(9);
\draw [->](I)--(s);
\draw [->](s)--(J);
\draw [->](s)--(4);
\draw [->](J)--(2);
\end{tikzpicture}
\qquad
\begin{tikzpicture}[thick,>=stealth,xscale=0.6,yscale=0.8]
\draw (4,0) node[tre] (r) {}; \etqb r 0
\draw (2,-1) node[tre] (c) {}; \etqb c 1
\draw (6,-1) node[tre] (n) {}; \etqb n 1
\draw (4,-2) node[hyb] (A) {}; \etqb A 1
\draw (0,-3) node[tre] (b) {}; \etqb b 2
\draw (9,-3) node[tre] (o) {}; \etqb o 2
\draw (1,-4) node[tre] (d) {}; \etqb d 4
\draw (6,-3) node[hyb] (B) {}; \etqb B 2
\draw (2,-5) node[tre] (f) {}; \etqb f 6
\draw (8,-5) node[tre] (e) {}; \etqb e 5
\draw (2,-6) node[tre] (h) {}; \etqb h 7
\draw (3,-6) node[hyb] (C) {}; \etqb C 6
\draw (6,-6) node[tre] (g) {}; \etqb g 6
\draw (3,-8) node[tre] (i) {}; \etqb i 7
\draw (7,-7) node[tre] (j) {}; \etqb j 7
\draw (2,-9.5) node[hyb] (D) {}; \etqb D 7
\draw (4,-7) node[hyb] (E) {}; \etqb E 7
\draw (4,-9) node[tre] (k) {}; \etqb k 8
\draw (6,-9) node[tre] (l) {}; \etqb l 8
\draw (7,-9) node[tre] (m) {}; \etqb m 8
\draw (0,-10) node[tre] (a) {}; \etqb a 3
\draw (5,-10) node[hyb] (F) {}; \etqb F 8
\draw (7,-10) node[hyb] (G) {}; \etqb G 8
\draw (9,-10) node[tre] (p) {}; \etqb p 3
\draw (8,-11) node[hyb] (H) {}; \etqb H 3
\draw (8,-12) node[tre] (q) {}; \etqb q 4
\draw (3,-12) node[hyb] (I) {}; \etqb I 4
\draw (3,-13) node[tre] (s) {}; \etqb s 5
\draw (1,-13) node[hyb] (J) {}; \etqb J 5
\draw (0,-14) node[tre] (1) {}; \etq 1
\draw (1,-14) node[tre] (2) {}; \etq 2 
\draw (2,-14) node[tre] (3) {}; \etq 3
\draw (3,-14) node[tre] (4) {}; \etq 4
\draw (4,-14) node[tre] (5) {}; \etq 5
\draw (5,-14) node[tre] (6) {}; \etq 6
\draw (6,-14) node[tre] (7) {}; \etq 7
\draw (7,-14) node[tre] (8) {}; \etq 8
\draw (8,-14) node[tre] (9) {}; \etq 9
\draw (9,-14) node[tre] (10) {}; \etq {10}
\draw [->](r)--(c);
\draw [->](r)--(n);
\draw [->](c)--(b);
\draw [->](c)--(A);
\draw [->](n)--(A);
\draw [->](n)--(o);
\draw [->](A)--(d);
\draw [->](b)--(a);
\draw [->](b)--(B);
\draw [->](o)--(B);
\draw [->](o)--(p);
\draw [->](d)--(f);
\draw [red,->](d)--(I);
\draw [->](B)--(e);
\draw [->](f)--(h);
\draw [->](f)--(C);
\draw [->](e)--(g);
\draw [red,->](e)--(J);
\draw [->](h)--(D);
\draw [->](h)--(E);
\draw [->](C)--(i);
\draw [->](g)--(C);
\draw [->](g)--(j);
\draw [->](i)--(D);
\draw [->](i)--(k);
\draw [->](j)--(E);
\draw [->](j)--(m);
\draw [->](D)--(3);
\draw [->](E)--(l);
\draw [->](k)--(5);
\draw [->](k)--(F);
\draw [->](l)--(7);
\draw [->](l)--(G);
\draw [->](m)--(F);
\draw [->](m)--(G);
\draw [->](F)--(6);
\draw [->](G)--(8);
\draw [->](a)--(1);
\draw [->](a)--(H);
\draw [->](p)--(H);
\draw [->](p)--(10);
\draw [->](H)--(q);
\draw [->](q)--(I);
\draw [->](q)--(9);
\draw [->](I)--(s);
\draw [->](s)--(J);
\draw [->](s)--(4);
\draw [->](J)--(2);
\end{tikzpicture}
\end{center}
\caption{\label{fig:contr3} The networks $N_{5}$ (left) and $N_{6}$
(right)}
\end{figure}
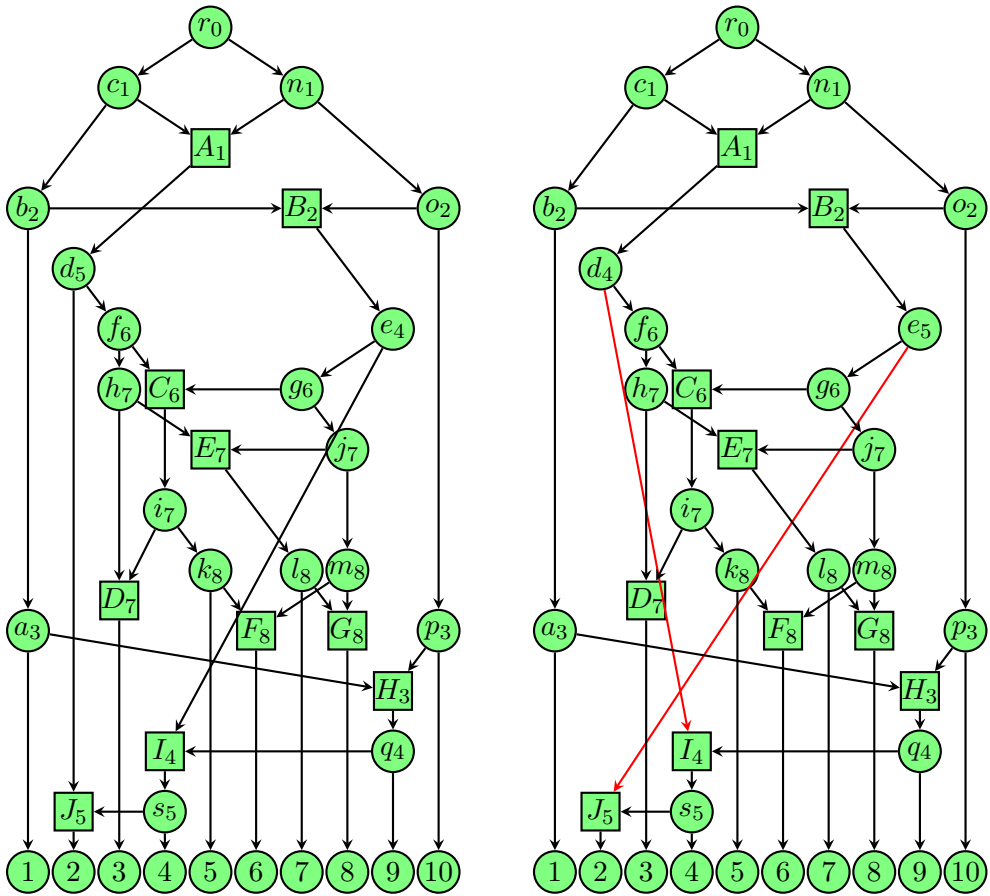

\begin{table}[htb]
{\scriptsize  \begin{tabular}{|c|c|c|c|c|}
     \hline
       \multicolumn{1}{|c}{arc's}& \multicolumn{2}{|c|}{$N_{5}$} & 
       \multicolumn{2}{|c|}{$N_{6}$}  \\
 \cline{2-5}
     head &\quad $A$\quad{} & $B$ &\quad $A$\quad{} & $B$   \\
      \hline
$a$ & 1 & $2_{3},4_{2},9_{1}$ & 1 & $2_{3},4_{2},9_{1}$ \\  
\hline
$b$ & 1 & $2_{3},3_{3},4_{2},5_{2},6_{3},7_{2},8_{3},9_{1}$ & 1 & $2_{3},3_{3},4_{2},5_{2},6_{3},7_{2},8_{3},9_{1}$ \\  
\hline
$c$ & 1 & $2_{3},3_{3},4_{2},5_{2},6_{3},7_{2},8_{3},9_{1}$ & 1 & $2_{3},3_{3},4_{2},5_{2},6_{3},7_{2},8_{3},9_{1}$ \\ 
\hline
$d$ & $\emptyset$ & $2_{1},3_{2},5_{1},6_{2},7_{1},8_{2}$ & $\emptyset$ & $2_{1},3_{2},4_{1},5_{1},6_{2},7_{1},8_{2}$ \\  
\hline
$e$ & $\emptyset$ & $2_{2},3_{2},4_{1},5_{1},6_{2},7_{1},8_{2}$ & $\emptyset$ &
$2_{2},3_{2},5_{1},6_{2},7_{1},8_{2}$ \\  
\hline
$f$ & $\emptyset$ & $3_{2},5_{1},6_{2},7_{1},8_{2}$ & $\emptyset$ & $3_{2},5_{1},6_{2},7_{1},8_{2}$ \\  
\hline
$g$ & $\emptyset$ & $3_{2},5_{1},6_{2},7_{1},8_{2}$ & $\emptyset$ & $3_{2},5_{1},6_{2},7_{1},8_{2}$ \\
\hline
$h$ & $\emptyset$ & $3_{1},7_{1},8_{2}$ & $\emptyset$ & $3_{1},7_{1},8_{2}$ \\  
\hline
$i$ & 5 & $3_{1},6_{1}$ & 5 & $3_{1},6_{1}$ \\  
\hline
$j$ &  $\emptyset$ & $6_{1},7_{1},8_{2}$ & $\emptyset$ & $6_{1},7_{1},8_{2}$ \\  
\hline
$k$ & 5 & $6_{1}$ & 5 & $6_{1}$ \\  
\hline
$l$ & 7 & $8_{1}$ & 7 & $8_{1}$ \\  
\hline
$m$ & $\emptyset$  & $6_{1},8_{1}$ & $\emptyset$  & $6_{1},8_{1}$ \\
\hline
$n$ & 10 & $2_{3},3_{3},4_{2},5_{2},6_{3},7_{2},8_{3},9_{1}$ & 10 & $2_{3},3_{3},4_{2},5_{2},6_{3},7_{2},8_{3},9_{1}$ \\  
\hline
$o$ & 10 & $2_{3},3_{3},4_{2},5_{2},6_{3},7_{2},8_{3},9_{1}$ & 10 & $2_{3},3_{3},4_{2},5_{2},6_{3},7_{2},8_{3},9_{1}$ \\  
\hline
$p$ & $10$ & $2_{3},4_{2},9_{1}$ & $10$ & $2_{3},4_{2},9_{1}$ \\  
\hline
$q$ & $9$ & $2_{2},4_{1}$ & $9$ & $2_{2},4_{1}$ \\
\hline
$s$ & 4 & $2_{1}$ & 4 & $2_{1}$ \\
\hline 
& $1,2_{3},3_{3},4_{2},$ & &
$1,2_{3},3_{3},4_{2},$ & \\
$r$  & $5_{2},6_{3},7_{2},$ & $\emptyset$ & $
5_{2},6_{3},7_{2},$ & $\emptyset$\\
& $8_{3},9_{1},10$ & & $8_{3},9_{1},10$ & \\
\hline
$A$ & $\emptyset$ & $2_{2},3_{3},5_{2},6_{3},7_{2},8_{3}$  &  $\emptyset$ &
$2_{2},3_{3},4_{2},5_{2},6_{3},7_{2},8_{3}$  \\
\hline
$B$ & $\emptyset$ & $2_{3},3_{3},4_{2},5_{2},6_{3},7_{2},8_{3}$  & $\emptyset$ & $2_{3},3_{3},5_{2},6_{3},7_{2},8_{3}$ \\
\hline
$C$ & $5_{1}$ & $3_{2},6_{2}$  & $5_{1}$ & $3_{2},6_{2}$   \\
\hline
$D$ & $3_{1}$ & $\emptyset$ & $3_{1}$ & $\emptyset$  \\
\hline
$E$ & $7_{1}$ & $8_{2}$ & $7_{1}$ & $8_{2}$ \\
\hline
$F$ & $6_{1}$ & $\emptyset$ & $6_{1}$ & $\emptyset$\\
\hline
$G$ & $8_{1}$ & $\emptyset$ & $8_{1}$ & $\emptyset$\\
\hline
$H$ & $9_{1}$ & $2_{3},4_{2}$ & $9_{1}$ & $2_{3},4_{2}$\\
\hline
$I$ & $4_{1}$ & $2_{2}$ & $4_{1}$ & $2_{2}$\\
\hline
$J$ & $2_{1}$ & $\emptyset$ & $2_{1}$ & $\emptyset$\\
\hline
 \end{tabular} 
}   \smallskip
   \caption{Tripartitions of the networks in Fig.~\ref{fig:contr3}}
\label{tbl:contr4}
\end{table}

\begin{table}[htb]
{\scriptsize  \begin{tabular}{|c|c|c|}
     \hline
       \multicolumn{1}{|c}{arc's}& \multicolumn{1}{|c|}{$N_{5}$} & 
       \multicolumn{1}{|c|}{$N_{6}$}  \\
 \cline{2-3}
     head & $RS$ & $RS$   \\
      \hline
$A$ &\ $\big\{\{1,2,\ldots,8,9\},\{2,3,\ldots,9,10\}\big\}$\ {} &\
$\big\{\{1,2,\ldots,8,9\},\{2,3,\ldots,9,10\}\big\}$\ {} \\
\hline
$B$ & $\big\{\{1,2,\ldots,8,9\},\{2,3,\ldots,9,10\}\big\}$ &
$\big\{\{1,2,\ldots,8,9\},\{2,3,\ldots,9,10\}\big\}$\\
\hline
$C$ & $\big\{\{3,5,6,7,8\},\{3,5,6,7,8\}\big\}$ & $\big\{\{3,5,6,7,8\},\{3,5,6,7,8\}\big\}$  \\
\hline
$D$ & $\big\{\{3,7,8\},\{3,5,6\}\big\}$ & $\big\{\{3,7,8\},\{3,5,6\}\big\}$  \\
\hline
$E$ & $\big\{\{3,7,8\},\{6,7,8\}\big\}$ & $\big\{\{3,7,8\},\{6,7,8\}\big\}$ \\
\hline
$F$ & $\big\{\{5,6\},\{6,8\}\big\}$ & $\big\{\{5,6\},\{6,8\}\big\}$ \\
\hline
$G$ & $\big\{\{7,8\},\{6,8\}\big\}$ &
$\big\{\{7,8\},\{6,8\}\big\}$\\
\hline
$H$ & $\big\{\{1,2,4,9\},\{2,4,9,10\}\big\}$ & $\big\{\{1,2,4,9\},\{2,4,9,10\}\big\}$ \\
\hline
$I$ & $\big\{\{2,3,4,5,6,7,8\},\{2,4,9\}\big\}$ & $\big\{\{2,3,4,5,6,7,8\},\{2,4,9\}\big\}$\\
\hline
$J$ & $\big\{\{2,4\},\{2,3,5,6,7,8\}\big\}$ & $\big\{\{2,4\},\{2,3,5,6,7,8\}\big\}$\\
\hline
 \end{tabular} 
} \smallskip
   \caption{Reticulation scenarios of the hybrid nodes of the networks in
   Fig.~\ref{fig:contr3}}  \label{tbl:contr5}
\end{table}

\begin{figure}[htb]
\begin{center}
\begin{tikzpicture}[thick,>=stealth,scale=0.9]
\draw (1,0) node[tre] (r) {}; \etqb r 0
\draw (0,-1) node[tre] (a) {}; \etqb a 1
\draw (2,-1) node[tre] (b) {}; \etqb b 1
\draw (1,-2) node[hyb] (A) {}; \etqb A 1
\draw (3,-2) node[tre] (c) {}; \etqb c 2
\draw (1,-3) node[tre] (d) {}; \etqb d 4
\draw (2,-3) node[tre] (e) {}; \etqb e 3
\draw (3,-3) node[hyb] (B) {}; \etqb B 3
\draw (4,-3) node[tre] (f) {}; \etqb f 3
\draw (3,-4) node[tre] (g) {}; \etqb g 4
\draw (1,-5) node[hyb] (C) {}; \etqb C 4
\draw (3,-5) node[hyb] (D) {}; \etqb D 4
\draw (0,-6) node[tre] (1) {}; \etq 1
\draw (1,-6) node[tre] (2) {}; \etq 2
\draw (2,-6) node[tre] (3) {}; \etq 3
\draw (3,-6) node[tre] (4) {}; \etq 4
\draw (4,-6) node[tre] (5) {}; \etq 5
\draw [->](r)--(a);
\draw [->](r)--(b);
\draw [->](a)--(1);
\draw [->](a)--(A);
\draw [->](b)--(A);
\draw [->](A)--(d);
\draw [->](b)--(c);
\draw [->](c)--(e);
\draw [->](c)--(f);
\draw [->](d)--(C);
\draw [->](d)--(D);
\draw [->](e)--(3);
\draw [->](e)--(B);
\draw [->](f)--(B);
\draw [->](f)--(5);
\draw [->](B)--(g);
\draw [->](g)--(C);
\draw [->](g)--(D);
\draw [->](C)--(2);
\draw [->](D)--(4);
\end{tikzpicture}
\qquad
\begin{tikzpicture}[thick,>=stealth,scale=0.9]
\draw (1,0) node[tre] (r) {}; \etqb r 0
\draw (0,-1) node[tre] (a) {}; \etqb a 1
\draw (2,-1) node[tre] (b) {}; \etqb b 1
\draw (1,-2) node[hyb] (A) {}; \etqb A 1
\draw (1,-3) node[tre] (d) {}; \etqb d 3
\draw (2,-3) node[tre] (e) {}; \etqb e 2
\draw (3,-3) node[hyb] (B) {}; \etqb B 2
\draw (4,-3) node[tre] (f) {}; \etqb f 2
\draw (3,-4) node[tre] (g) {}; \etqb g 3
\draw (1,-5) node[hyb] (C) {}; \etqb C 3
\draw (3,-5) node[hyb] (D) {}; \etqb D 3
\draw (0,-6) node[tre] (1) {}; \etq 1
\draw (1,-6) node[tre] (2) {}; \etq 2
\draw (2,-6) node[tre] (3) {}; \etq 3
\draw (3,-6) node[tre] (4) {}; \etq 4
\draw (4,-6) node[tre] (5) {}; \etq 5
\draw [->](r)--(a);
\draw [->](r)--(b);
\draw [->](a)--(1);
\draw [->](a)--(A);
\draw [->](b)--(A);
\draw [->](A)--(d);
\draw [->](b)--(e);
\draw [->](b)--(f);
\draw [->](d)--(C);
\draw [->](d)--(D);
\draw [->](e)--(3);
\draw [->](e)--(B);
\draw [->](f)--(B);
\draw [->](f)--(5);
\draw [->](B)--(g);
\draw [->](g)--(C);
\draw [->](g)--(D);
\draw [->](C)--(2);
\draw [->](D)--(4);
\end{tikzpicture}
\end{center}
\caption{\label{fig:contr4} The networks $N_{7}$ (left) and $N_{8}$
(right)}
\end{figure}
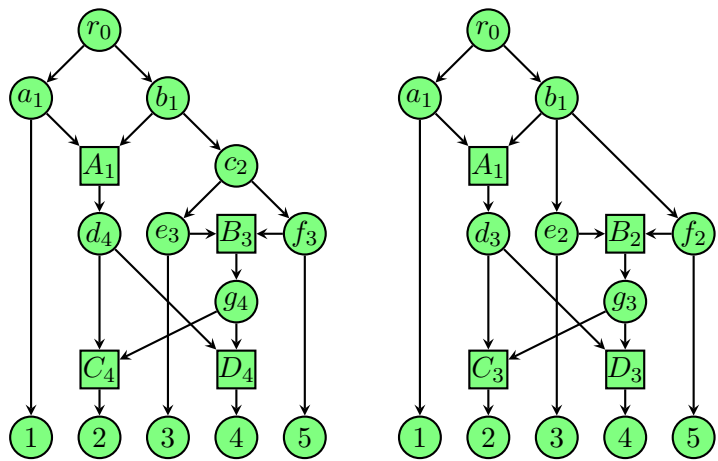

\begin{figure}[htb]
\begin{center}
\begin{tikzpicture}[thick,>=stealth]
\draw (2,0) node[tre] (r) {}; \etq r
\draw (0,-2) node[tre] (a) {}; \etq a
\draw (1,-4) node[hyb] (U) {}; 
\draw (3,-4) node[hyb] (V) {}; 
\draw (0,-6) node[tre] (1) {}; \etq 1
\draw (1,-6) node[tre] (2) {}; \etq 2
\draw (2,-6) node[tre] (3) {}; \etq 3
\draw (3,-6) node[tre] (4) {}; \etq 4
\draw (4,-6) node[tre] (5) {}; \etq 5
\draw [->](r)--(a);
\draw [->](r)--(U);
\draw [->](r)--(V);
\draw [->](r)--(3);
\draw [->](r)--(5);
\draw [->](U)--(2);
\draw [->](V)--(4);
\draw [->](a)--(1);
\draw [->](a)--(U);
\draw [->](a)--(V);
\end{tikzpicture}
\qquad
\begin{tikzpicture}[thick,>=stealth]
\draw (2,0) node[tre] (r) {}; \etq r
\draw (0,-1.5) node[tre] (a) {}; \etq a
\draw (3,-1.5) node[tre] (b) {}; \etq b
\draw (2,-2) node[tre] (e) {}; \etq e
\draw (4,-3) node[tre] (f) {}; \etq f
\draw (1,-4.5) node[hyb] (U) {}; 
\draw (3,-4.5) node[hyb] (V) {}; 
\draw (0,-6) node[tre] (1) {}; \etq 1
\draw (1,-6) node[tre] (2) {}; \etq 2
\draw (2,-6) node[tre] (3) {}; \etq 3
\draw (3,-6) node[tre] (4) {}; \etq 4
\draw (4,-6) node[tre] (5) {}; \etq 5
\draw [->](r)--(a);
\draw [->](r)--(b);
\draw [->](a)--(1);
\draw [->](a)--(U);
\draw [->](a)--(V);
\draw [->](a)--(1);
\draw [->](b)--(U);
\draw [->](b)--(e);
\draw [->](b)--(f);
\draw [->](b)--(V);
\draw [->](U)--(2);
\draw [->](V)--(4);
\draw [->](e)--(U);
\draw [->](e)--(3);
\draw [->](f)--(V);
\draw [->](f)--(5);
\end{tikzpicture}
\end{center}
\caption{\label{fig:contr4.2} The reduced versions $R(N_{7})$
(left) and $R(N_{8})$ (right) of the networks given in Fig.~\ref{fig:contr4}}
\end{figure}
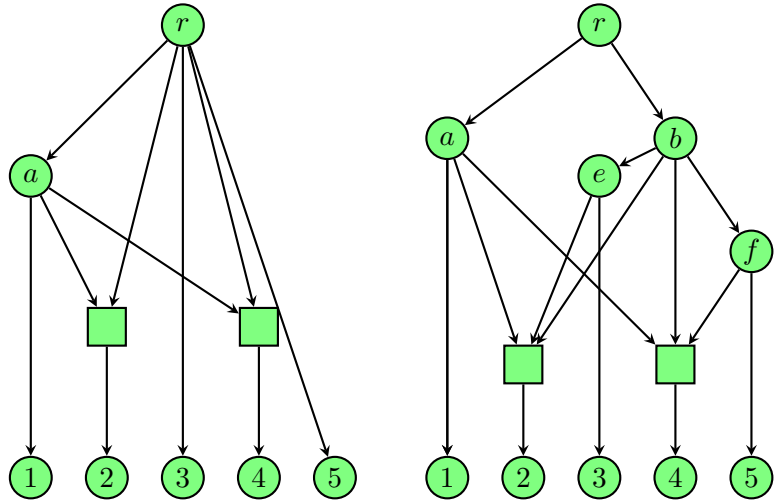

\begin{table}[htb]
{\scriptsize  \begin{tabular}{|c|c|c|c|c|}
     \hline
       \multicolumn{1}{|c}{arc's}& \multicolumn{2}{|c|}{$N_{7}$} & 
       \multicolumn{2}{|c|}{$N_{8}$}  \\
 \cline{2-5}
     head &\quad $A$\quad{} & $B$ &\quad $A$\quad{} & $B$   \\
      \hline
$a$ & 1 & $2_{2},4_{2}$ & 1 & $2_{2},4_{2}$ \\  
\hline
$b$ & $3,5$ & $2_{2},4_{2}$ & $3,5$  & $2_{2},4_{2}$ \\  
\hline
$c$ & $3,5$ & $2_{2},4_{2}$ & --  & -- \\  
\hline
$d$ & $\emptyset$ & $2_{1},4_{1}$ & $\emptyset$ & $2_{1},4_{1}$\\
\hline
$e$ & $3$ & $2_{2},4_{2}$ &  $3$ & $2_{2},4_{2}$	\\	
\hline
$f$ & $5$ & $2_{2},4_{2}$ & $5$ & $2_{2},4_{2}$\\
\hline
$g$ & $\emptyset$ & $2_{1},4_{1}$ & $\emptyset$ & $2_{1},4_{1}$\\
\hline
$r$ & $1,2_{2},3,4_{2},5$ & $\emptyset$ &$1,2_{2},3,4_{2},5$ & $\emptyset$\\
\hline
$A$& $\emptyset$ & $2_{2},4_{2}$ & $\emptyset$ & $2_{2},4_{2}$\\
\hline
$B$ & $\emptyset$ & $2_{2},4_{2}$ & $\emptyset$ & $2_{2},4_{2}$ \\
\hline
$C$ & $2_{1}$ & $\emptyset$ & $2_{1}$ & $\emptyset$\\
\hline
$D$ & $4_{1}$ & $\emptyset$ & $4_{1}$ & $\emptyset$\\
\hline
\end{tabular} 
}   \smallskip
   \caption{Tripartitions of the networks in Fig.~\ref{fig:contr4}}
\label{tbl:contr6}
\end{table}

\begin{figure}[htb]
\begin{center}
\begin{tikzpicture}[thick,>=stealth,scale=0.9]
\draw (2,1) node[tre] (r) {}; \etqb r 0
\draw (0,0) node[tre] (a) {}; \etqb a 1
\draw (2,-0.5) node[tre] (b) {}; \etqb b 1
\draw (4,0) node[tre] (c) {}; \etqb c 1
\draw (1,-2) node[hyb] (A) {}; \etqb A 1
\draw (3,-2) node[hyb] (B) {}; \etqb B 1
\draw (0,-3) node[tre] (1) {}; \etq 1
\draw (1,-3) node[tre] (2) {}; \etq 2
\draw (3,-3) node[tre] (3) {}; \etq 3
\draw (4,-3) node[tre] (4) {}; \etq 4
\draw [->](r)--(a);
\draw [->](r)--(b);
\draw [->](r)--(c);
\draw [->](a)--(1);
\draw [->](a)--(A);
\draw [->](a)--(B);
\draw [->](b)--(A);
\draw [->](b)--(B);
\draw [->](c)--(A);
\draw [->](c)--(B);
\draw [->](c)--(4);
\draw [->](A)--(2);
\draw [->](B)--(3);
\end{tikzpicture}
\qquad
\begin{tikzpicture}[thick,>=stealth,scale=0.9]
\draw (2,0) node[tre] (r) {}; \etqb r 0
\draw (0,-1) node[tre] (a) {}; \etqb a 2
\draw (2,-1.5) node[tre] (b) {}; \etqb b 2
\draw (4,-1) node[tre] (c) {}; \etqb c 1
\draw (1,-3) node[hyb] (A) {}; \etqb A 2
\draw (3,-3) node[hyb] (B) {}; \etqb B 2
\draw (0,-4) node[tre] (1) {}; \etq 1
\draw (1,-4) node[tre] (2) {}; \etq 2
\draw (3,-4) node[tre] (3) {}; \etq 3
\draw (4,-4) node[tre] (4) {}; \etq 4
\draw [->](r)--(a);
\draw [->](r)--(c);
\draw [->](a)--(1);
\draw [->](a)--(A);
\draw [->](a)--(B);
\draw [->](b)--(A);
\draw [->](b)--(B);
\draw [->](c)--(b);
\draw [->](c)--(4);
\draw [->](A)--(2);
\draw [->](B)--(3);
\end{tikzpicture}
\end{center}
\caption{\label{fig:contr6} The networks $N_{9}$ (left) and $N_{10}$
(right)}
\end{figure}
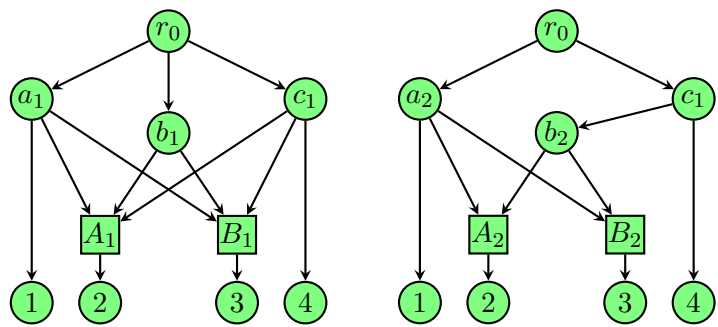

\begin{table}[htb]
{\scriptsize  \begin{tabular}{|c|c|c|c|c|}
     \hline
       \multicolumn{1}{|c}{arc's}& \multicolumn{2}{|c|}{$N_{9}$} & 
       \multicolumn{2}{|c|}{$N_{10}$}  \\
 \cline{2-5}
     head &\quad $A$\quad{} & $B$ &\quad $A$\quad{} & $B$   \\
      \hline
$a$ & 1 & $2_{1},3_{1}$ & 1 & $2_{1},3_{1}$ \\
\hline
$b$ & $\emptyset$ &  $2_{1},3_{1}$ & $\emptyset$ &  $2_{1},3_{1}$ \\
\hline
$c$ & 4 &  $2_{1},3_{1}$  & 4 &  $2_{1},3_{1}$\\
\hline
$r$ & $1,2_{1},3_{1},4$ & $\emptyset$ &  $1,2_{1},3_{1},4$ & $\emptyset$
\\
\hline
$A$ & $2_{1}$ & $\emptyset$ &$2_{1}$ & $\emptyset$
\\
\hline
$B$ & $3_{1}$ & $\emptyset$ &$3_{1}$ & $\emptyset$
\\
\hline
\end{tabular} 
}   \smallskip
  \caption{Tripartitions of the networks in Fig.~\ref{fig:contr6}}
 \label{tbl:contr7.2}
\end{table}

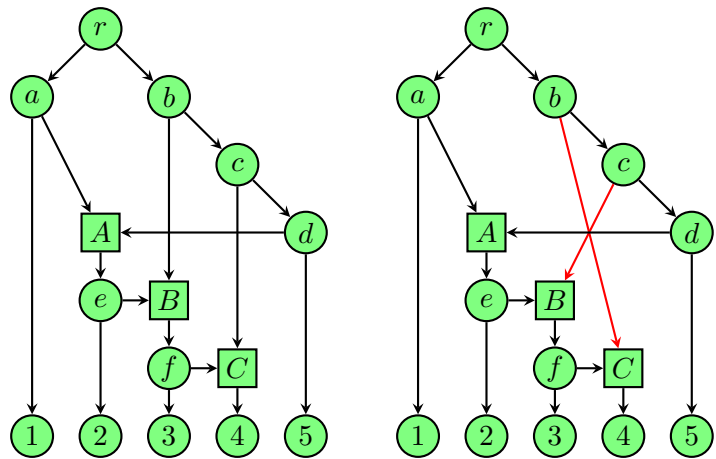
\begin{figure}[htb]
\begin{center}
\begin{tikzpicture}[thick,>=stealth,scale=0.9]
\draw (3,7) node[tre] (r) {}; \etq r
\draw (2,6) node[tre] (a) {}; \etq a
\draw (4,6) node[tre] (b) {}; \etq b
\draw (2,1) node[tre] (1) {}; \etq 1
\draw (5,5) node[tre] (c) {}; \etq c
\draw (3,4) node[hyb] (A) {}; \etq A
\draw (6,4) node[tre] (d) {}; \etq d
\draw (3,3) node[tre] (e) {}; \etq e
\draw (4,3) node[hyb] (B) {}; \etq B
\draw (6,1) node[tre] (5) {}; \etq 5
\draw (3,1) node[tre] (2) {}; \etq 2
\draw (4,2) node[tre] (f) {}; \etq f
\draw (5,2) node[hyb] (C) {}; \etq C
\draw (4,1) node[tre] (3) {}; \etq 3
\draw (5,1) node[tre] (4) {}; \etq 4
\draw [->](r)--(a);
\draw [->](r)--(b);
\draw [->](a)--(1);
\draw [->](a)--(A);
\draw [->](b)--(B);
\draw [->](b)--(c);
\draw [->](c)--(C);
\draw [->](c)--(d);
\draw [->](A)--(e);
\draw [->](d)--(A);
\draw [->](d)--(5);
\draw [->](e)--(2);
\draw [->](e)--(B);
\draw [->](B)--(f);
\draw [->](f)--(3);
\draw [->](f)--(C);
\draw [->](C)--(4);
\end{tikzpicture}
\qquad
\begin{tikzpicture}[thick,>=stealth,scale=0.9]
\draw (3,7) node[tre] (r) {}; \etq r
\draw (2,6) node[tre] (a) {}; \etq a
\draw (4,6) node[tre] (b) {}; \etq b
\draw (2,1) node[tre] (1) {}; \etq 1
\draw (5,5) node[tre] (c) {}; \etq c
\draw (3,4) node[hyb] (A) {}; \etq A
\draw (6,4) node[tre] (d) {}; \etq d
\draw (3,3) node[tre] (e) {}; \etq e
\draw (4,3) node[hyb] (B) {}; \etq B
\draw (6,1) node[tre] (5) {}; \etq 5
\draw (3,1) node[tre] (2) {}; \etq 2
\draw (4,2) node[tre] (f) {}; \etq f
\draw (5,2) node[hyb] (C) {}; \etq C
\draw (4,1) node[tre] (3) {}; \etq 3
\draw (5,1) node[tre] (4) {}; \etq 4
\draw [->](r)--(a);
\draw [->](r)--(b);
\draw [->](a)--(1);
\draw [->](a)--(A);
\draw [red,->](b)--(C);
\draw [->](b)--(c);
\draw [red,->](c)--(B);
\draw [->](c)--(d);
\draw [->](A)--(e);
\draw [->](d)--(A);
\draw [->](d)--(5);
\draw [->](e)--(2);
\draw [->](e)--(B);
\draw [->](B)--(f);
\draw [->](f)--(3);
\draw [->](f)--(C);
\draw [->](C)--(4);
\end{tikzpicture}
\end{center}
\caption{\label{fig:treech-nocomp} The networks $N_{11}$ (left) and
$N_{12}$ (right) are tree-child but cannot be distinguished by means
of their tripartitions}
\end{figure}

\begin{table}[htb]
{\scriptsize  \begin{tabular}{|c|c|c|c|c|}
     \hline
\multicolumn{1}{|c}{arc's}& \multicolumn{2}{|c}{$N_{11}$} & 
\multicolumn{2}{|c|}{$N_{12}$}  \\
  \cline{2-5}
      head &\quad $A$\quad{} & $B$  & \quad $A$\quad{} & $B$    \\
      \hline
      $a$ & $1$ & $2_{1},3_{2},4_{3}$  & $1$ & $2_{1},3_{2},4_{3}$ 
      \\
      \hline
      $b$ & $5$ & $2_{1},3_{2},4_{3}$  & $5$ & $2_{1},3_{2},4_{3}$  
      \\
      \hline
      $c$ & $5$ & $2_{1},3_{2},4_{3}$  & $5$ & $2_{1},3_{2},4_{3}$  
      \\
      \hline
      $d$ & $5$ & $2_{1},3_{2},4_{3}$  & $5$ & $2_{1},3_{2},4_{3}$  
      \\
      \hline
      $e$ & $2$ & $3_{1},4_{2}$  & $2$ & $3_{1},4_{2}$  
      \\
      \hline
      $f$ & $3$ & $4_{1}$ & $3$ & $4_{1}$  \\
      \hline
       $r$ & $1,2_{1},3_{2},4_{3},5$ & $\emptyset$ & $1,2_{1},3_{2},4_{3},5$ & $\emptyset$ \\
       \hline
       $A$ & $2_{1}$ & $3_{2},4_{3}$   & $2_{1}$ & $3_{2},4_{3}$   \\
         \hline
       $B$ & $3_{1}$ & $4_{2}$   &  $3_{1}$ & $4_{2}$     \\
      \hline
      $C$ & $4_{1}$ & $\emptyset$   &  $4_{1}$ &
      $\emptyset$     \\
\hline
 \multicolumn{1}{|c}{ }& \multicolumn{2}{|c}{$RS$} & 
\multicolumn{2}{|c|}{$RS$}  \\
\hline
     \multicolumn{1}{|c}{ $A$} & \multicolumn{2}{|c}{$\big\{\{1,2,3,4\},\{2,3,4,5\}\big\}$} & 
     \multicolumn{2}{|c|}{$\big\{\{1,2,3,4\},\{2,3,4,5\}\big\}$}  \\
         \hline
       \multicolumn{1}{|c}{$B$} & \multicolumn{2}{|c}{$\big\{\{2,3,4\},\{2,3,4,5\}\big\}$} &  \multicolumn{2}{|c|}{$\big\{\{2,3,4\},\{2,3,4,5\}\big\}$}   \\
      \hline
        \multicolumn{1}{|c}{$C$} & \multicolumn{2}{|c}{$\big\{\{3,4\},\{2,3,4,5\}\big\}$} &  \multicolumn{2}{|c|}{$\big\{\{3,4\},\{2,3,4,5\}\big\}$}   \\
 \hline
 \end{tabular} 
}\smallskip
     \caption{Tripartitions and reticulation scenarios of the
     tree-child phylogenetic networks in Fig.~\ref{fig:treech-nocomp}}
   \label{tbl:treech-nocomp}
\end{table}

\end{document}